\documentclass[a4paper,english,fleqn,leqno,orivec]{llncs}
\makeatletter\let\endnote\@undefined\makeatother
\usepackage[utf8]{inputenc}
\usepackage{babel}
\useshorthands{"}
\defineshorthand{"=}{\babelhyphen{hard}}
\defineshorthand{"~}{\babelhyphen{nobreak}}
\usepackage{newtxtext}
\usepackage{amssymb,stmaryrd,mathrsfs,mathtools}
\usepackage{microtype}
\usepackage{xspace,xifthen,xstring,calc}
\usepackage[hidelinks,raiselinks,bookmarks=false]{hyperref}
\usepackage{csquotes}
\usepackage[capitalise,nameinlink]{cleveref}
\crefname{section}{Sect.}{Sects.}\Crefname{section}{Section}{Sections}
\crefname{figure}{Fig.}{Figs.}\Crefname{figure}{Figure}{Figures}
\crefname{table}{Tab.}{Tabs.}\Crefname{table}{Table}{Tables}
\crefname{appendix}{App.}{Apps.}\Crefname{appendix}{Appendix}{Appendices}
\crefformat{equation}{(#2#1#3)}\crefrangeformat{equation}{(#3#1#4--#5#2#6)}
\crefname{enumi}{}{}\Crefname{enumi}{}{}
\crefformat{enumi}{(#2#1#3)}\crefrangeformat{enumi}{(#3#1#4--#5#2#6)}
\crefname{definition}{Def.}{Defs.}\Crefname{definition}{Definition}{Definitions}
\crefname{lemma}{Lem.}{Lems.}\Crefname{lemma}{Lemma}{Lemmata}
\crefname{proposition}{Prop.}{Props.}\Crefname{proposition}{Proposition}{Propositions}
\crefname{theorem}{Thm.}{Thms.}\Crefname{theorem}{Theorem}{Theorems}
\crefname{corollary}{Cor.}{Cors.}\Crefname{corollary}{Corollary}{Corollaries}
\crefname{example}{Ex.}{Exs.}\Crefname{example}{Example}{Examples}
\usepackage{enumitem}
\setlist[itemize]{leftmargin=*, label={--}}
\setlist[enumerate]{leftmargin=*}
\newlist{refpars}{enumerate}{1}
\setlist[refpars]{label={(\alph*)}, ref={\alph*}, leftmargin=0pt, labelindent=0pt, labelwidth=*, align=left, topsep=4pt plus 4pt, itemsep=4pt plus 2pt, partopsep=0pt, listparindent=\parindent}
\crefname{refparsi}{}{}\Crefname{refparsi}{}{}
\crefformat{refparsi}{(#2#1#3)}\crefrangeformat{refparsi}{(#3#1#4--#5#2#6)}
\usepackage{tikz}
\usetikzlibrary{automata,decorations.pathmorphing,positioning,matrix,shapes,arrows,calc}
\usepackage{booktabs,diagbox,xfrac}
\usepackage{thmtools}

\usepackage{pdfcomment}
\newif\ifshowednotes\showednotestrue
\newcommand*{\ednoteauthor}{EdNote}
\newcommand*{\ednotecomment}{No comment.}
\newcommand*{\myenotezwritemark}[1]{\leavevmode\marginpar{\pdftooltip{\footnotesize\ednoteauthor(#1)}{\ednotecomment}}}
\usepackage{enotez}
\setenotez{backref=true}
\setenotez{list-name={EdNotes}}
\setenotez{mark-cs={\myenotezwritemark}}
\newcommand{\ednote}[2][Ednote]{%
  \ifshowednotes%
    \renewcommand*{\ednoteauthor}{#1}%
    \renewcommand*{\ednotecomment}{#2}%
    \bgroup%
      \fontsize{6pt}{6pt}\selectfont%
      \endnote{\ifthenelse{\equal{#1}{Ednote}}{#2}{#1: #2}}%
    \egroup%
  \fi%
}
\ifshowednotes%
  \AtEndDocument{\printendnotes}%
\fi

\allowdisplaybreaks

\newcommand*{\powerset}{\mathnormal{\wp}}

\newcommand*{\restrictop}{\mathnormal{\upharpoonright}}
\newcommand*{\restrict}[2]{#1\restrictop#2}
\newcommand*{\limp}{\mathbin{\rightarrow}}
\newcommand*{\lequiv}{\mathbin{\leftrightarrow}}

\newcommand*{\truefrm}{\mathrm{true}}
\newcommand*{\falsefrm}{\mathrm{false}}
\newcommand*{\sem}[1]{\llbracket#1\rrbracket}

\newcommand*{\pll}{\mathbin{\|}}

\newcommand*{\bisimrel}{\mathrel{\approx}}

\newcommand*{\modelequiv}{\models\joinrel\mathrel|}

\newcommand*{\trans}[2][M]{\xrightarrow{#2}\ifthenelse{\isempty{#1}}{}{_{#1}}}

\mathchardef\cln\mathcode`\:
\mathcode`\:=\string"8000
\begingroup \catcode`\:=\active
  \gdef:{\nobreak\mskip2mu\mathpunct{}\nonscript
    \mkern-\thinmuskip{\cln}\mskip6muplus1mu\relax}
  \endgroup
\newcommand*{\dclnrel}{\mathrel{\cln\cln}}

\NewDocumentCommand{\stacked}{O{t}O{1}m}{\bgroup\renewcommand{\arraystretch}{#2}\begin{array}[#1]{@{}l@{}}#3\end{array}\egroup}
\newcommand{\structrule}[2]{%
  \dfrac{%
    \renewcommand{\arraystretch}{1.1}%
    \begin{array}{@{}c@{}}#1\end{array}%
  }{%
    \renewcommand{\arraystretch}{1.1}%
    \begin{array}{@{}l@{}}#2\end{array}%
  }%
}

\newcommand*{\nof}[1]{\text{\ensuremath{#1}}}

\NewDocumentCommand{\IfNoValuesTF}{ooomm}{%
  \IfNoValueTF{#1}{%
    \IfNoValueTF{#2}{%
      \IfNoValueTF{#3}{%
        #4%
      }{%
        #5%
      }
    }{%
      #5%
    }%
  }{%
    #5%
  }%
}     

\newcommand{\ie}{i.\kern1pt e.\relax\xspace}
\newcommand{\eg}{e.\kern1pt g.\relax\xspace}
\newcommand{\wrt}{w.\kern1pt r.\kern1pt t.\relax\xspace}
\makeatletter
\renewcommand{\xRightarrow}[2][]{\ext@arrow 0359\Rightarrowfill@{#1}{#2}}
\makeatother

\NewDocumentCommand{\bcl}{}{\mathit{bcl}}
\NewDocumentCommand{\befrm}{}{\phi}
\NewDocumentCommand{\wlpop}{e{_}}{\mathrm{wlp}_{\mkern-1.5mu\ESig_{#1}}}
\NewDocumentCommand{\Wlpop}{e{_}}{\mathrm{Wlp}_{\mkern-1.5mu\ESig_{#1}}}
\NewDocumentCommand{\wlp}{e{_}mm}{\wlpop_{#1}(#2, #3)}
\NewDocumentCommand{\Wlp}{e{_}mm}{\Wlpop_{#1}(#2, #3)}

\NewDocumentCommand{\Prop}{e{_}}{\Pi\IfValueT{#1}{_{\mathit{#1}}}}
\NewDocumentCommand{\Ag}{e{_}}{A\IfValueT{#1}{_{\mathit{#1}}}}
\NewDocumentCommand{\ESig}{e{_}}{\Sigma\IfValueT{#1}{_{\mathit{#1}}}}
\NewDocumentCommand{\prop}{o}{\IfNoValueTF{#1}{p}{\mathrm{#1}}}
\NewDocumentCommand{\ag}{so}{\IfBooleanTF{#1}{\IfNoValueTF{#2}{a}{#2}}{\IfNoValueTF{#2}{\nof{a}}{\mathrm{#2}}}}
\NewDocumentCommand{\agfrm}{m}{\IfBooleanTF{#1}{\ag*}{\ag}}
\NewDocumentCommand{\agG}{so}{\IfBooleanTF{#1}{\IfNoValueTF{#2}{G}{#2}}{\IfNoValueTF{#2}{\nof{G}}{\mathrm{#2}}}}
\NewDocumentCommand{\agGfrm}{m}{\IfBooleanTF{#1}{\agG*}{\agG}}

\NewDocumentCommand{\estr}{}{K}
\NewDocumentCommand{\eacc}{}{E}
\NewDocumentCommand{\ewrlds}{}{W}
\NewDocumentCommand{\ewrld}{}{w}
\NewDocumentCommand{\elab}{}{L}
\NewDocumentCommand{\ESt}{e{_}}{\mathcal{E}_{\ESig_{#1}}}
\newcommand*{\est}{\mathfrak{K}}
\NewDocumentCommand{\ESts}{e{_}}{\mathcal{S}_{\ESig_{#1}}}
\newcommand*{\ests}{\mathcal{K}}
\NewDocumentCommand{\estsfam}{e{|}}{\vec{\ests}\IfValueT{#1}{\restrictop{#1}}}
\NewDocumentCommand{\dests}{se{|.}}{%
  \IfNoValueTF{#3}{%
    \estsfam|{#2}
  }{%
    \IfNoValueTF{#2}{%
      \ests_{\agfrm{#1}[#3]}%
    }{%
      (\estsfam|{#2})_{\agfrm{#1}[#3]}%
    }%
  }%
}
\NewDocumentCommand{\emodels}{e{_}}{\models_{\ESig_{#1}}}
\NewDocumentCommand{\emodelequiv}{e{_}}{\modelequiv_{\ESig_{#1}}}
\NewDocumentCommand{\esem}{e{_}m}{\sem{#2}_{\ESig_{#1}}}
\NewDocumentCommand{\ebisim}{e{_}}{\bisimrel_{\ESig_{#1}}}

\NewDocumentCommand{\ags}{e{_}}{\mathrm{ags}_{\mkern-1.5mu\ESig_{#1}}}
\NewDocumentCommand{\EFrmset}{e{_}}{\mathscr{F}_{\ESig_{#1}}}
\NewDocumentCommand{\EFrm}{se{|_}}{%
  \EFrmset_{#3}\IfValueT{#2}{\restrictop\agfrm{#1}[#2]}%
}
\NewDocumentCommand{\efrm}{}{\varphi}
\NewDocumentCommand{\efrms}{}{\Phi}
\NewDocumentCommand{\Kop}{}{\mathsf{K}}
\NewDocumentCommand{\K}{se{_}m}{\mathnormal{\Kop}_{\agfrm{#1}[#2]}\!\mathop{}#3}
\NewDocumentCommand{\Mop}{}{\mathsf{M}}
\NewDocumentCommand{\M}{se{_}m}{\mathnormal{\Mop}_{\agfrm{#1}[#2]}\!\mathop{}#3}

\NewDocumentCommand{\eastr}{}{U}
\NewDocumentCommand{\eapnts}{}{Q}
\newcommand*{\eapnt}[1][]{\ifthenelse{\isempty{#1}}{\text{\ensuremath{q}}}{\mathsf{#1}}}
\NewDocumentCommand{\eaacc}{}{F}
\NewDocumentCommand{\eapre}{}{\mathit{pre}}
\NewDocumentCommand{\eaPre}{}{\mathit{Pre}}
\NewDocumentCommand{\EActset}{e{_}}{\mathcal{U}_{\ESig_{#1}}}
\NewDocumentCommand{\EAct}{se{_|}}{%
  \EActset_{#2}\IfValueT{#3}{\restrictop\agfrm{#1}[#3]}%
}
\newcommand*{\eact}{\mathfrak{u}}
\NewDocumentCommand{\eupd}{e{_}}{\mathbin{\lhd}_{\ESig_{#1}}}

\NewDocumentCommand{\DEFrm}{e{_}}{\mathscr{F}^{[]}_{\EESig_{#1}}}

\newcommand*{\dlbox}[2]{[#1]#2}
\newcommand*{\dldia}[2]{\langle#1\rangle#2}

\NewDocumentCommand{\ecact}{}{\alpha}
\NewDocumentCommand{\ECActset}{e{_}}{\mathcal{A}_{\ESig_{#1}}}
\NewDocumentCommand{\ECAct}{se{_|}}{\IfNoValueTF{#3}{\ECActset_{#2}}{\restrict{\ECActset_{#2}}{\agfrm{#1}[#3]}}}

\NewDocumentCommand{\EEFrm}{e{_}}{\mathscr{D}_{\EESig_{#1}}}
\NewDocumentCommand{\eefrm}{}{\psi}
\NewDocumentCommand{\eemodels}{e{_}}{\models_{\EESig_{#1}}^{\eecact_{#1}}}

\NewDocumentCommand{\Fcsset}{t{^}e{_}}{\IfBooleanT{#1}{\vec}\Phi\IfValueT{#2}{_{\mathit{#2}}}}
\NewDocumentCommand{\Fcs}{se{_|}}{\Fcsset_{#2}\IfValueT{#3}{\restrictop\agfrm{#1}[#3]}}
\NewDocumentCommand{\Fcsfam}{e{_}}{\Fcsset^_{#1}}
\NewDocumentCommand{\SESig}{oe{_}}{\IfNoValueTF{#2}{(\ESig,\allowbreak \IfNoValueTF{#1}{\Fcs}{#1})}{(\ESig_{#2},\allowbreak \IfNoValueTF{#1}{\Fcs_{#2}}{#1})}}

\NewDocumentCommand{\SESt}{e{_}}{\mathcal{S}_{\ESig_{#1}}^{\Fcsset_{#1}}}
\NewDocumentCommand{\sestset}{}{\Gamma}
\NewDocumentCommand{\sest}{se{|}}{\sestset\IfValueT{#2}{\restrictop\agfrm{#1}[#2]}}
\NewDocumentCommand{\sesem}{e{_}m}{\sem{#2}_{\ESig_{#1}}^{\Fcsset_{#1}}}
\NewDocumentCommand{\semodels}{oe{_|}}{\models_{\ESig_{#2}}^{\IfNoValueTF{#1}{\Fcs_{#2}|{#3}}{#1}}}

\NewDocumentCommand{\SEActset}{e{_}O{\Fcs}}{\mathcal{U}_{\ESig_{#1}}^{#2_{#1}}}
\NewDocumentCommand{\SEAct}{se{_}O{\Fcs}e{|}}{\SEActset_{#2}[#3]\IfValueT{#4}{\restrictop\agfrm{#1}[#4]}}
\NewDocumentCommand{\seupd}{oe{_|}}{\eupd_{#2}^{\IfNoValueTF{#1}{\Fcs_{#2}|{#3}}{#1}}}
\NewDocumentCommand{\sseupd}{oe{_|}}{\mathbin{\ll}_{\ESig_{#2}}^{\IfNoValueTF{#1}{\Fcs_{#2}|{#3}}{#1}}}
\NewDocumentCommand{\erepr}{}{\rho}
\NewDocumentCommand{\SECActset}{e{_}}{\mathcal{A}_{\ESig_{#1}}^{\Fcsset_{#1}}}
\NewDocumentCommand{\SECAct}{se{_|}}{\IfNoValueTF{#3}{\SECActset_{#2}}{\restrict{\ECActset_{#2}}{\agfrm{#1}[#3]}}}

\NewDocumentCommand{\SEEFrm}{e{_}}{\mathscr{D}_{\EESig_{#1}}^{\Fcs_{#1}}}

\NewDocumentCommand{\seequiv}{e{_}}{\equiv_{\ESig_{#1}}^{\Fcs_{#1}}}

\newcommand*{\grpann}{\mathit{grp}}
\newcommand*{\sndlos}[3]{\mathit{snd}_{\text{los}}^{\ag[#1]\to\ag[#2]}(#3)}%
\newcommand*{\sndrel}[3]{\mathit{snd}_{\text{rel}}^{\ag[#1]\to\ag[#2]}(#3)}%
\newcommand*{\comgrp}[3][]{\ifthenelse{\isempty{#1}}{\grpann}{\grpann^{\eapnt[#1]}}_{\ag[#2]}(#3)}%

\NewDocumentCommand{\EEActSym}{e{_.}}{\mathrm{N}\IfValueTF{#1}{_{\mathit{#1}\IfValueT{#2}{,\ag[#2]}}}{\IfValueT{#2}{_{\ag[#2]}}}}
\newcommand*{\eeactsym}{\eta}
\NewDocumentCommand{\EESig}{e{_}}{\check{\Sigma}\IfValueT{#1}{_{\mathit{#1}}}}
\NewDocumentCommand{\eeags}{e{_}}{\mathrm{ags}_{\mkern-1.5mu\EESig_{#1}}}
\NewDocumentCommand{\eeag}{e{_}}{\mathrm{ag}_{\mkern-1.5mu\EESig_{#1}}}
\NewDocumentCommand{\eecact}{e{_}}{\mathit{act}\IfValueT{#1}{_{\mkern-1.5mu\mathit{#1}}}}

\NewDocumentCommand{\EEProcset}{e{_}}{\mathcal{P}_{\EESig_{#1}}}
\NewDocumentCommand{\EEProc}{se{_|}}{\EEProcset_{#2}\IfValueT{#3}{\restrictop\agfrm{#1}[#3]}}
\NewDocumentCommand{\eecnd}{}{\varphi}
\NewDocumentCommand{\eecnds}{}{\widehat{\varphi}}
\NewDocumentCommand{\eeproc}{}{P}
\NewDocumentCommand{\eenil}{}{\mathbf{0}}
\NewDocumentCommand{\eempty}{}{\epsilon}

\NewDocumentCommand{\eens}{}{\vec{E}}
\NewDocumentCommand{\eensfam}{E{_}{{\ag \in \Ag}}m}{(#2)_{#1}}
\NewDocumentCommand{\epctrans}{e{_}m}{\mathrel{\xhookrightarrow{#2}{}\mkern-5mu_{\EESig_{#1}}}}
\NewDocumentCommand{\eectrans}{e{_}m}{\mathrel{\xhookrightarrow{#2}{}\mkern-5mu_{\EESig_{#1}}}}
\NewDocumentCommand{\eetrans}{e{_}m}{\mathrel{\xrightarrow{#2}{}\mkern-5mu_{\EESig_{#1}}^{\eecact_{#1}}}}

\NewDocumentCommand{\EEActset}{e{_}}{\mathcal{C}_{\EESig_{#1}}}
\NewDocumentCommand{\EEAct}{se{_|}}{\IfNoValueTF{#3}{\EEActset_{#2}}{\restrict{\EEActset_{#2}}{\agfrm{#1}[#3]}}}
\NewDocumentCommand{\eeact}{}{\lambda}
\NewDocumentCommand{\eesem}{e{_}m}{\sem{#2}_{\EESig_{#1}}^{\eecact_{#1}}}
\NewDocumentCommand{\ewits}{}{\sigma}

\NewDocumentCommand{\seetrans}{e{_}m}{\mathrel{\xrightarrow{#2}\mkern-5mu{}_{\EESig_{#1}, \Fcsset_{#1}}^{\eecact_{#1}}}}
\NewDocumentCommand{\seesem}{e{_}m}{\sem{#2}_{\EESig_{#1}, \Fcs_{#1}}^{\eecact_{#1}}}
\NewDocumentCommand{\seemodels}{e{_}}{\models_{\EESig_{#1}, \Fcs_{#1}}^{\eecact_{#1}}}

\NewDocumentCommand{\beecnds}{}{\widehat{\phi}}
\NewDocumentCommand{\SEEAct}{e{_}}{\EEActset_{#1}^{\Fcs_{#1}}}

\NewDocumentCommand{\seeequiv}{e{_}}{\equiv_{\EESig_{#1}, \Fcs_{#1}}^{\eecact_{#1}}}

\newif\ifarXiv
\arXivtrue

\title{\ifarXiv Epistemic Ensembles in Semantic and Symbolic\\ Environments (Extended Version with Proofs)\else
                Epistemic Ensembles in\\ Semantic and Symbolic Environments\fi}
\author{%
  Rolf Hennicker\inst{1} \and
  Alexander Knapp\inst{2} \and
  Martin Wirsing\inst{1}
}
\institute{
  Ludwig-Maximilians-Universität München, München, Germany\\
 \email{$\{$hennicker$,\;$wirsing$\}$@ifi.lmu.de}
\and 
  Universität Augsburg, Augsburg, Germany\\
  \email{knapp@informatik.uni-augsburg.de}\\[2ex]
  \emph{Dedicated to Rocco De Nicola} 
}

\begin{document}

\maketitle

\begin{abstract}
An epistemic ensemble is composed of knowledge"=based agents capable of
retrieving and sharing knowledge and beliefs about themselves and their peers.
These agents access a global knowledge state and use actions to communicate and
cooperate, altering the collective knowledge state. We study two types of
mathematical semantics for epistemic ensembles based on a common syntactic
operational ensemble semantics: a semantic environment defined by a class of
global epistemic states, and a symbolic environment consisting of a set of
epistemic formulæ.  For relating these environments, we use the concept of
$\Fcs$"=equivalence, where a class of epistemic states and a knowledge base are
$\Fcs$"=equivalent, if any formula of $\Fcs$ holds in the class of epistemic
states if, and only if, it is an element of the knowledge base.  Our main
theorem shows that $\Fcs$"=equivalent configurations simulate each other and
satisfy the same dynamic epistemic ensemble formulae.
\end{abstract}

\section{Introduction}\label{sec:intro}

Ensembles \cite{Hölzl_Rauschmayer_Wirsing:SIS:2008,ascens15} are collective
systems consisting of dynamically interacting autonomic entities.  In an
epistemic ensemble~\cite{hennicker-knapp-wirsing:isola:2022} these entities are
epistemic agents that possess knowledge and beliefs about themselves and other
agents.  The agents have access to a global knowledge state and utilise
epistemic actions to cooperate and communicate. An epistemic action involves
announcing an agent’s knowledge or belief to others, thereby altering the
collective knowledge state.

This paper builds upon our previous work
\cite{hennicker-knapp-wirsing:isola:2022,Hennicker_Knapp_Wirsing:LPAR:2024}. In
\cite{hennicker-knapp-wirsing:isola:2022}, we introduced epistemic
ensembles in a semantic environment consisting of a single epistemic state; in

\cite{Hennicker_Knapp_Wirsing:LPAR:2024}, we considered epistemic processes
and related them in semantic environments of a single epistemic state with
symbolic environments.  Here, we focus on ensembles and model them as
families of concurrently running epistemic processes.  We introduce a syntactic
operational semantics for these epistemic ensembles.  Building on the
operational semantics we consider an extended notion of semantic environment
consisting not only of a single epistemic state but of a class of epistemic
states.  Following ideas from~\cite{Hennicker_Knapp_Wirsing:LPAR:2024}, we also
apply the symbolic approach to epistemic ensembles and relate symbolic
environments with the new, more general semantic approach.

The dynamic behaviour of each agent process in an ensemble is specified using a simple
process algebra with guards and recursion.  The syntactic operational semantics
of the entire ensemble is defined generically by conditional transitions over
uninterpreted agent actions.  To specify global properties of epistemic
ensembles, we use propositional dynamic logic over compound ensemble actions.
Our main result is the study of two complementary kinds of mathematical
semantics for epistemic ensembles building on the common syntactic operational
semantics: one in a semantic environment defined by a class of global epistemic
states, represented by pointed Kripke structures, and the other in a symbolic
environment defined by a global epistemic knowledge base, represented by a
finite set of epistemic formulæ.  The agent actions are first interpreted by
pointed action
models~\cite{baltag-moss-solecki:tark:1998,baltag-renne:stanford:2016}; in the
semantic environment their effect is given by product updates on Kripke
structures~\cite{baltag-moss-solecki:tark:1998,baltag-renne:stanford:2016},
while in the symbolic environment we use weakest liberal
preconditions~\cite{dijkstra-scholten:1990}.  Building
on~\cite{Hennicker_Knapp_Wirsing:LPAR:2024}, we relate the semantic and symbolic
environments by fixing a finite set $\Fcs$ of focus formulæ: A class $\ests$ of
epistemic states is $\Fcs$"=equivalent to a symbolic state $\sest \subseteq
\Fcs$ if all $\est \in \ests$ satisfy exactly those $\efrm \in \Fcs$ that are in
$\sest$.  Reusing the notion of $\Fcs$"=representability we ensure that semantic
and symbolic updates preserve $\Fcs$"=equivalence.  As our main results we show
that $\Fcs$"=equivalent ensemble configurations simulate each other and that
they satisfy the same dynamic ensemble formulæ.

\paragraph{Related work.}
Research on using epistemic logic in system modelling and programming began with
the seminal works on Dynamic Epistemic Logic
(DEL~\cite{baltag-moss-solecki:tark:1998,van-ditmarsch-van-der-hoek-kooi:2008})
and knowledge-based programs~\cite{fagin-et-al:2003}.  DEL focuses on modelling
and verifying knowledge changes induced by action execution.  We use DEL's
reduction rules to define the weakest liberal precondition, which also underpin
many foundational results of DEL, such as soundness and completeness
\cite{baltag-moss-solecki:tark:1998,baltag-renne:stanford:2016}, as well as the
definition of a sequent calculus \cite{Wirsing_Knapp:IGPL:2023}. Knowledge-based
programs consider systems of concurrently running agents. 
\cite{DBLP:journals/jphil/BenthemGHP09} examines protocols based on DEL and,
as~\cite{parikh-ramanujam:jlli:2003}, system properties in epistemic temporal
logic.

Our work is based on a process"=oriented description of dynamic system
behaviour~\cite{DeNicola2011}, in alignment with modern
languages for ensembles such as SCEL~\cite{DeNicola_et_al:ASCENS:2015},
CARMA~\cite{Bortolussi_et_al:QAPL:2015}, and DEECo
\cite{DBLP:conf/cbse/BuresGHKKP13}. However, these languages rely on
communication mechanisms like message passing and predicate-based communication,
rather than epistemic actions.  Our symbolic semantics for epistemic
representation is related to the syntactic structures in
\cite{hennicker-knapp-wirsing:isola:2022} and belief bases in
\cite{DBLP:conf/kr/LoriniPS22}. While Lorini's work is also based on DEL, it
lacks a process-algebraic setting and uses specialised forms of knowledge update
operations. In contrast, we consider general action models and apply an
abstraction technique from epistemic Kripke semantics to a symbolic approach,
preserving and reflecting dynamic ensemble properties.

\paragraph{Structure of the paper.}
We first introduce epistemic ensembles and their syntactic operational semantics
in \cref{sec:eensembles} and dynamic ensemble formulæ in \cref{sec:elogic}.  The
agent actions of epistemic ensembles are interpreted by action models in
\cref{sec:action-models}.  \Cref{sec:eens-semantic} and \cref{sec:eens-symbolic}
present the semantic and the symbolic environment of epistemic ensembles.  The
equivalence of both semantics is shown in \cref{sec:eens-semantic-symbolic}.  We
conclude in \cref{sec:conclusions}.\ifarXiv\else\footnote{An extended version with all proofs is available at \url{https://arxiv.org}.}\fi

\section{Epistemic Ensembles}\label{sec:eensembles}

An epistemic ensemble is formed by a collection of agents which run concurrently
to accomplish a certain task. In the epistemic context collaboration of agents
is achieved by the execution of agent actions where agents inform other agents
about (parts of) their knowledge, which may concern themselves, other agents or
the environment.  Each agent follows a certain protocol which is given by an
epistemic process description.  In this section we introduce the syntactic
notions for building epistemic ensembles and we provide a generic operational
semantics for epistemic ensembles which will be instantiated later on for
executing ensembles in semantic and symbolic environments.

\paragraph{Epistemic formulæ.}
Epistemic formulæ provide the means to describe knowledge; see, e.g.,
\cite{baltag-renne:stanford:2016,van-ditmarsch-van-der-hoek-kooi:2008,fagin-et-al:2003}.
An \emph{epistemic signature} $\ESig = (\Prop, \Ag)$ consists of a set $\Prop$
of (atomic) \emph{propositions} and a set $\Ag$ of \emph{agents}.
The set $\EFrm$ of \emph{epistemic formulæ} $\efrm$ 
over $\ESig = (\Prop, \Ag)$ is defined by the following grammar:
\begin{equation*}
  \efrm
{\;\cln\cln=\;}
  \prop \;\mid\;
  \truefrm \;\mid\;
  \neg\efrm \;\mid\;
  \efrm_1 \land \efrm_2\;\mid\;
  \K{\efrm}
\quad\text{where $\prop \in \Prop$ and $\ag \in \Ag$.}
\end{equation*}
The epistemic formula $\K{\efrm}$ is to be read as ``agent $\ag$ \emph{knows
  $\varphi$}''.  We use the usual Boolean shorthand notations like $\falsefrm$
for $\neg\truefrm$, $\varphi_1 \lor \varphi_2$ for $\neg(\neg\varphi_1 \land
\neg\varphi_2)$, etc.  Moreover, we write $\M{\efrm}$ for $\neg\K{\neg\efrm}$;
this latter epistemic modality is dual to $\K{}$ and to be read as ``agent $\ag$
\emph{deems $\efrm$ possible}''.  For each $\ag \in \Ag$, the set $\EFrm|\ag$ of
\emph{$\ag$"=epistemic formulæ} $ \efrm_{\ag}$ having an $\ag$"=modality as
their top operator, is given by the following grammar:
\begin{equation*}
  \efrm_{\ag}
{\;\cln\cln=\;}
  \truefrm \;\mid\;
  \neg\efrm_{\ag} \;\mid\;
  \efrm_{\ag, 1} \land \efrm_{\ag, 2}\;\mid\;
  \K{\efrm}
\quad\text{where $\efrm \in \EFrm$.}
\end{equation*}
For any $\varphi \in \EFrm$ the set of
possible \emph{agents} of $\varphi$ is given by
$\ags(\varphi) = \{ \ag \in \Ag \mid \varphi \in \EFrm|\ag \}$.
In our setting, $\ags(\varphi)$ is either a singleton or $\Ag$
(if no $\K{}$ occurs in $\varphi$).

\begin{example}\label{ex:witzel-ens-1}
Our running example is inspired by~\cite{witzel-zvesper:aamas:2008}.
We consider a set of two agents $\Ag_2 = \{ \ag[1], \ag[2] \}$ each one holding a bit
$\prop[x]_i \in \Prop_2 = \{ \prop[x_1], \prop[x_2] \}$.
The epistemic signature is $\ESig_2 = (\Prop_2, \Ag_2)$.
In a situation where $\prop[x_1]$ is true, the $\ag[1]$"=formula
$\K_{\ag[1]}{\prop[x_1]}$ expresses that agent $\ag[1]$ knows this.
The $\ag[2]$"=formula 
$\K_{\ag[2]}{\prop[x_1]} \lor \K_{\ag[2]}{\neg\prop[x_1]}$ says that agent
$\ag[2]$  knows the value of $\prop[x_1]$ but we cannot infer its concrete
value. The $\ag[1]$"=formula
$\neg\K_{\ag[1]}(\K_{\ag[2]}{\prop[x_1]} \lor \K_{\ag[2]}{\neg\prop[x_1]})$
expresses that agent $\ag[1]$ does not know whether agent $\ag[2]$
knows the value of $\prop[x_1]$.
\end{example}

\paragraph{Epistemic ensemble signatures.}
Our formalisation of epistemic ensembles is based on the notion of an
\emph{epistemic ensemble signature} $\EESig = (\ESig, (\EEActSym.\ag)_{\ag \in
  \Ag})$ consisting of an epistemic signature $\ESig = (\Prop, \Ag)$ and an
$\Ag$"=family $(\EEActSym.\ag)_{\ag \in \Ag}$ of pairwise disjoint sets
$\EEActSym.\ag$ of \emph{agent action symbols}, briefly called \emph{agent
  actions}.  Each set $\EEActSym.\ag$ determines which actions are possible for
agent $\ag$.  We write $\bigcup\EEActSym$ for $\bigcup_{\ag \in \Ag}
\EEActSym.\ag$ and for each $\eeactsym \in \bigcup\EEActSym$ we set
$\eeags(\eeactsym) = \{\ag\}$ if $\eeactsym \in \EEActSym.\ag$.

\begin{example}\label{ex:witzel-ens-2}
Continuing~\cref{ex:witzel-ens-1} we introduce an action $\mathit{stop}$ and two
actions $\mathit{tell}12(\prop[x_1])$ and $\mathit{tell}12(\neg\prop[x_1])$ for
agent $\ag[1]$, \ie, $\EEActSym_2.1 = \{\mathit{stop},
\mathit{tell}12(\prop[x_1]), \mathit{tell}12(\neg\prop[x_1]) \}$, the latter two
to express that agent $\ag[1]$ tells the value of $\prop[x_1]$ to agent $\ag[2]$
but, in each case, the information transfer is unreliable since agent $\ag[2]$
may be too far away.  On the other hand, agent $\ag[2]$ (when it got the value
of $\prop[x_1]$) may acknowledge in a reliable way the reception with the action
$\mathit{ack}21(x_1)$, \ie, $\EEActSym_2.2 = \{ \mathit{ack}21(x_1) \}$.
\end{example}

\smallskip\noindent%
\textbf{General assumption.}  In the sequel, we always assume given an epistemic
ensemble signature $\EESig = (\ESig, (\EEActSym.\ag)_{\ag \in \Ag})$ with
underlying epistemic signature $\ESig = (\Prop, \Ag)$.

\paragraph{Epistemic processes.}
We consider an epistemic process language for describing the behaviour of agents
which participate in an epistemic ensemble.  The set $\EEProc$ of
\emph{epistemic processes} $\eeproc$ over $\EESig$ is defined by the grammar
\begin{equation*}
  \eeproc
{\;\cln\cln=\;}\begin{array}[t]{@{}l@{}}
  \eenil \;\mid\;
  \eeactsym.\eeproc \;\mid\;
  \eecnd \supset \eeproc \;\mid\;
  \eeproc_1 + \eeproc_2\;\mid\;
  \mu X \,.\, \eeproc \;\mid\;
  X
\end{array}
\end{equation*}
where $\eenil$ represents the inactive process, $\eeactsym.\eeproc$ prefixes $\eeproc$
with an agent action $\eeactsym \in \bigcup\EEActSym$, $\eecnd \supset \eeproc$ is a
guarded process with condition $\eecnd \in \EFrm$, $\eeproc_1 + \eeproc_2$
denotes the non"=deterministic choice between processes $\eeproc_1$ and 
$\eeproc_2$, $\mu X \,.\, \eeproc$ is
a recursive process, and $X$ is a process variable typically used in recursive process definitions.

The \emph{operational semantics} of epistemic processes is given by
\emph{conditional transitions} $\eeproc \epctrans{\eecnds \cln \eeactsym}
\eeproc'$ relating a process $\eeproc$ via a guard $\eecnds \in \EFrm$ and
an agent action $\eeactsym \in \bigcup\EEActSym$ with another process
$\eeproc'$.  The transitions are defined inductively by the rules in
\cref{tab:proc-rules}, where
successive guards are conjoined and $\truefrm$
represents the empty guard.
\begin{table}[t!]\centering
$\begin{array}{@{}c@{\quad}l@{\qquad}c@{}}
\eeactsym.\eeproc \epctrans{\truefrm \cln \eeactsym} \eeproc
& &
\structrule{\eeproc \epctrans{\eecnds \cln \eeactsym} \eeproc'}{\eecnd \supset \eeproc \epctrans{\eecnds \land \eecnd \cln \eeactsym} \eeproc'}
\\[4ex]
\structrule{\eeproc_{\ell} \epctrans{\eecnds \cln \eeactsym} \eeproc_{\ell}'}{\eeproc_1 + \eeproc_2 \epctrans{\eecnds \cln \eeactsym} \eeproc_{\ell}'}
& \text{for }\ell \in \{ 1, 2 \}
&
\structrule{\eeproc\{ X \mapsto \mu X \,.\, \eeproc \} \epctrans{\eecnds \cln \eeactsym} \eeproc'}{\mu X \,.\, \eeproc \epctrans{\eecnds \cln \eeactsym} \eeproc'}
\end{array}$\\[1.5ex]
\caption{Rules for epistemic processes}\label{tab:proc-rules}
\end{table}

Epistemic processes are used to describe the behaviour of
single agents. But not any process expression in $\EEProc$ is meaningful for any agent.
For instance, a process $\eeactsym.\eenil$ can be carried out by an agent $\ag$
only if the action symbol $\eeactsym$ represents an action of $\ag$,
\ie, $\eeactsym \in \EEActSym.\ag$.
For any process $\eeproc \in \EEProc$ the set $\eeags(\eeproc)$ of agents that are allowed to execute  $\eeproc$ is inductively defined by
\begin{equation*}
\renewcommand{\arraystretch}{1.2}\begin{array}{@{}l@{\quad}l@{}}
\eeags(\eenil) = \Ag &
\eeags(\eeactsym.\eeproc) = \eeags(\eeactsym) \cap \eeags(\eeproc)\\
\eeags(\eecnd \supset \eeproc) = \ags(\eecnd) \cap \eeags(\eeproc)&
\eeags(\eeproc_1 + \eeproc_2) = \eeags(\eeproc_1) \cap \eeags(\eeproc_2)\\
\eeags(\mu X \,.\, \eeproc) = \eeags(\eeproc)&
\eeags(X) = \Ag
\end{array}
\end{equation*}
For executing a process $\eeproc \in \EEProc$ the following
kind of \emph{subject reduction} holds:

\begin{restatable}{lemma}{LemProcAgs}\label{lem:proc-ags}
If $\eeproc \epctrans{\eecnds \cln \eeactsym} \eeproc'$, then $\eeags(\eeproc)
\subseteq \ags(\eecnds) \cap \eeags(\eeactsym) \cap \eeags(\eeproc')$.
\end{restatable}

\paragraph{Epistemic ensembles.}
Now we have all ingredients to define epistemic ensembles
as families of epistemic agent processes.
Formally, an \emph{epistemic ensemble} over $\EESig$ is given by a family
$\eens = \eensfam{\ag \cln \eeproc_{\ag}}$ such that,
for each $\ag \in \Ag$, $\eeproc_{\ag}$ is an epistemic process in $\EEProc$ with
$\ag \in \eeags(\eeproc_{\ag})$, \ie,
$\ag$ is allowed to perform $\eeproc_{\ag}$.

For notational reasons and for defining the operational semantics of ensembles,
we also consider (sub-)families $\eensfam_{\ag
  \in \agG}{\ag \cln \eeproc_{\ag}}$ of agent processes with $\agG \subseteq
\Ag$; 
their
\emph{composition} $\eensfam_{\ag_1 \in \agG_1}{\ag_1 \cln \eeproc_{\ag_1}} \pll
\eensfam_{\ag_2 \in \agG_2}{\ag_2 \cln \eeproc_{\ag_2}}$ for disjoint $\agG_1,
\agG_2 \subseteq \Ag$ as the family $\eensfam_{\ag \in \agG_1 \cup \agG_2}{\ag
  \cln \eeproc_{\ag}}$; and, for $\ag*[i] \in \Ag$, the \emph{singleton}
$\ag*[i] \cln \eeproc_{\ag*[i]}$ which stands for $\eensfam_{\ag \in \{ \ag*[i]
  \}}{\ag \cln \eeproc_{\ag}}$.

\begin{example}\label{ex:witzel-ens}
Relying on the epistemic ensemble signature developed in~\cref{ex:witzel-ens-1}
and~\cref{ex:witzel-ens-2}, we consider the following simple epistemic ensemble
$\mathit{Sys}$ with two processes $\mathit{Ag1}$ for agent $\ag[1]$ and
$\mathit{Ag2}$ for agent $\ag[2]$.  In the process descriptions we abbreviate,
for each $a \in \{\ag[1],\ag[2]\}$, the formula $\K{\prop[x_1]} \lor
\K{\neg\prop[x_1]}$ by $\K{x_1}$ ($x_1$ written in italics).
\begin{align*}
\mathit{Ag1}
&=
\mu X \,.\, \big(\stacked{\neg\K_1{\K_2{x_1}} \supset (\stacked{\K_1{\prop[x_1]} \supset \mathit{tell}12(\prop[x_1]).X +{}\\ \K_1{\neg\prop[x_1]} \supset \mathit{tell}12(\neg\prop[x_1]).X) +{}}\\ \K_1{\K_2{x_1}} \supset \mathit{stop}.\eenil\big)}
\\
\mathit{Ag2}
&=
\K_2{x_1} \supset \mathit{ack}21(x_1).\eenil
\\
\mathit{Sys}
&= \ag[1] \cln \mathit{Ag1} \pll \ag[2] \cln \mathit{Ag2}
\end{align*}
By $\mathit{tell}12(\prop[x_1])$ and $\mathit{tell}12(\neg\prop[x_1])$ agent
$\ag[1]$ repeatedly ``tells'' agent $\ag[2]$ the value of $\prop[x_1]$ (in an
unreliable way) until it is sure that agent $\ag[2]$ knows the value; when
having indeed learnt the value of $\prop[x_1]$, agent $\ag[2]$ acknowledges this
fact (in a reliable way) to agent $\ag[1]$ with $\mathit{ack}21(x_1)$.
\end{example}

The \emph{syntactic operational semantics} of epistemic ensembles is given by
\emph{conditional transitions} $\eectrans{}$ relating an ensemble $\eens$ via a
guard $\eecnds \in \EFrm$ and an agent action $\eeactsym$ with another ensemble
$\eens'$ according to the following rule which is based on the rules for
processes:
\begin{equation*}\label{eq:sem-ens}\tag{$*$}
\ag*[i] \cln \eeproc_{\ag*[i]} \pll \eens \eectrans{\eecnds \cln \eeactsym} \ag*[i] \cln \eeproc_{\ag*[i]}' \pll \eens
\quad\text{if }\eeproc_{\ag*[i]} \epctrans{\eecnds \cln \eeactsym} \eeproc_{\ag*[i]}'
\end{equation*}
Note that each ensemble step is well"=defined, since $\ag*[i] \in
\eeags(\eeproc_{\ag*[i]})$ implies $\ag*[i] \in \eeags(\eeproc_{\ag*[i]}')$ by
\cref{lem:proc-ags}, and thus $\ag*[i] \cln \eeproc_{\ag*[i]}' \pll \eens$
is again an ensemble. Moreover, the ordering of processes in an ensemble is irrelevant since they are families, \ie, functions mapping agents to processes.

The rule \cref{eq:sem-ens} is formulated in a generic way without considering
evaluations of guards and effects of actions.  It provides, however, a
convenient basis for concrete instantiations.

\begin{example}\label{ex:witzel-ens-trans}
\Cref{fig:witzel-ens-trans} depicts the conditional transitions of the
bit transmission protocol in \cref{ex:witzel-ens}.
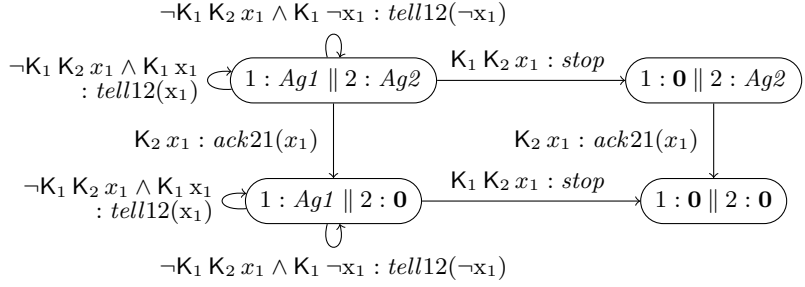
\begin{figure}[t!]\centering
\begin{tikzpicture}[%
  auto, scale=1, every loop/.style={-latex'},%
  font={\fontsize{9pt}{9pt}\selectfont},%
  every label/.append style={outer sep=0pt, inner sep=2pt, label distance=4pt},%
]
\tikzstyle{state}=[rounded rectangle, draw,inner sep=4pt,minimum width=1.0cm]
\tikzstyle{trans}=[draw,-latex']
\tikzstyle{annotation}=[]
\node[state] (s00) {$\ag[1] \cln \mathit{Ag1} \pll \ag[2] \cln \mathit{Ag2}$};
\node[state] (s10) at ($ (s00) + (5, 0) $) {$\ag[1] \cln \eenil \pll \ag[2] \cln \mathit{Ag2}$};
\node[state] (s01) at ($ (s00) + (0, -1.6) $) {$\ag[1] \cln \mathit{Ag1} \pll \ag[2] \cln \eenil$};
\node[state] (s11) at ($ (s10) + (0, -1.6) $) {$\ag[1] \cln \eenil \pll \ag[2] \cln \eenil$};
\path[trans, ->]
(s00) edge[loop, out=185, in=175, distance=12pt] node[left, align=right] {%
  $\neg\K_1{\K_2{x_1}} \land \K_1{\prop[x_1]}$\\%
  ${} \cln \mathit{tell}12(\prop[x_1])$%
} (s00)
(s00) edge[loop, out=105, in=75, distance=12pt] node[above, align=center] {%
  $\neg\K_1{\K_2{x_1}} \land \K_1{\neg\prop[x_1]} \cln \mathit{tell}12(\neg\prop[x_1])$%
} (s00)
(s00) edge node[above, align=left] {%
  $\K_1{\K_2{x_1}} \cln \mathit{stop}$
} (s10)
(s00) edge node[left, align=right] {%
  $\K_2{x_1} \cln \mathit{ack}21(x_1)$
} (s01)
(s01) edge[loop, out=185, in=175, distance=12pt] node[left, align=right] {%
  $\neg\K_1{\K_2{x_1}} \land \K_1{\prop[x_1]}$\\%
  ${} \cln \mathit{tell}12(\prop[x_1])$%
} (s01)
(s01) edge[loop, out=-105, in=-75, distance=12pt] node[below, align=center] {%
  $\neg\K_1{\K_2{x_1}} \land \K_1{\neg\prop[x_1]} \cln \mathit{tell}12(\neg\prop[x_1])$%
} (s01)
(s10) edge node[left, align=right] {%
  $\K_2{x_1} \cln \mathit{ack}21(x_1)$
} (s11)
(s01) edge node[above, align=left] {%
  $\K_1{\K_2{x_1}} \cln \mathit{stop}$
} (s11)
;
\end{tikzpicture}
\vspace*{-1.5ex}
\caption{Transition system for the syntactic bit transmission ensemble.}\label{fig:witzel-ens-trans}
\end{figure}
\end{example}

\section{Dynamic Epistemic Ensemble Logic}\label{sec:elogic}

So far we have considered a constructive approach for representing epistemic
ensembles with local processes for each agent.  We are now interested in a
declarative language for expressing and specifying behavioural properties of
epistemic ensembles from a global perspective.  For this purpose we use formulæ
in the style of propositional dynamic logic where regular expressions of agent
actions, called compound ensemble actions, are used as modalities. In contrast
to local agent processes, we have now a global view where actions of different
agents can be combined.

The set $\EEAct$ of \emph{compound ensemble actions} over $\EESig$ is defined by
the following grammar:
\begin{align*}
  \eeact
&{\;\cln\cln=\;}\begin{array}[t]{@{}l@{}}
  \eeactsym \;\mid\;
  \eecnd{?} \;\mid\;
  \eeact_1 + \eeact_2 \;\mid\;
  \eeact_1 ; \eeact_2 \;\mid\;
  \eeact^*
\quad\text{where $\eeactsym \in \bigcup\EEActSym$ and $\eecnd \in \EFrm$.}
  \end{array}
\end{align*}
Besides agent actions the compound ensemble actions include a \emph{test}
$\eecnd{?}$ on an epistemic formula $\eecnd \in \EFrm$, \emph{non"=deterministic choices}
$\eeact_1 + \eeact_2$ of compound ensemble actions $\eeact_1$ and $\eeact_2$, \emph{sequential compositions} $\eeact_1; \eeact_2$, and
\emph{sequential loops} $\eeact^*$.

Following the style of propositional dynamic logic the set $\EEFrm$ of
\emph{epistemic ensemble formulæ} over $\EESig$ is defined by the following
grammar where compound ensemble actions are used as modalities:
\begin{equation*}
  \eefrm
{\;\cln\cln=\;}
  \truefrm \;\mid\;
  \efrm \;\mid\;
  \neg\eefrm \;\mid\;
  \eefrm_1 \land \eefrm_2 \;\mid\;
  \dlbox{\eeact}{\eefrm}
\quad
\text{where $\efrm \in \EFrm$ and $\eeact \in \EEAct$.}
\end{equation*}
The formula $\dlbox{\eeact}{\eefrm}$ is to be read as ``after all possible
executions of the compound action $\eeact$ formula $\eefrm$ holds''.  We use the
usual abbreviations like $\falsefrm$ or $\lor$ as before, and we write
$\dldia{\eeact}{\eefrm}$ for $\neg\dlbox{\eeact}{\neg\eefrm}$; this latter
dynamic modality is dual to $\dlbox{\eeact}{}$ and to be read as ``there is some
execution of $\eeact$ such that $\eefrm$ holds afterwards''.

\begin{example}\label{ex:elogic}
For our two agent system we are interested in the following properties, in which
we abbreviate the compound action
$\mathit{stop} + \mathit{tell}12(\prop[x_1]) + \mathit{tell}12(\neg\prop[x_1]) + \mathit{ack}21(x_1)$ by ``$\mathit{some}$'':
 
\begin{enumerate}
  \item ``As long as agent $\ag[1]$ does not know that agent $\ag[2]$ knows the
value of $\prop[x_1]$ the ensemble will not stop'':\quad
$\dlbox{\mathit{some}^*}{\neg\K_1{\K_2{x_1}}} \limp
\dldia{\mathit{some}}{\truefrm}$

  \item ``Whenever agent $\ag[1]$ tells the value of $\prop[x_1]$ to agent
$\ag[2]$, it is possible that agent $\ag[2]$ will eventually know the
value'':\quad $\dlbox{\mathit{some}^*; (\mathit{tell}12(\prop[x_1]) +
  \mathit{tell}12(\neg\prop[x_1]))}{\dldia{\mathit{some}^*}{\K_2{x_1}}}$
\end{enumerate}
\end{example}
 
We cannot provide a formal satisfaction relation here saying when an epistemic
ensemble satisfies an epistemic ensemble formula, since we cannot talk yet about
satisfaction of ensemble formulæ as long as we do not have a concrete
formalisation of epistemic states and of the effect of agent actions on them.
But we can already now provide a definition to capture, for each compound
ensemble action $\eeact$, which sequences $\ewits = (\eecnds_1 \cln
\eeactsym_1^{\eempty}) \ldots (\eecnds_k \cln \eeactsym_k^{\eempty})$ of guard
and agent action pairs can be executed when moving stepwise from an ensemble
$\eens$ to an ensemble $\eens'$ in accordance with rule \cref{eq:sem-ens} from
above. 
Thereby concatenation of such two sequences $\ewits_1$ and $\ewits_2$ is denoted
by $\ewits_1 \cdot \ewits_2$ and $\eempty$ denotes an artificial empty agent
action.  The inductive rules in \cref{tab:ens-actions-rules} define when a
(stepwise) move $(\eens, \ewits, \eens')$ is a \emph{witness} for a compound
ensemble action $\eeact$, written $(\eens, \ewits, \eens') \dclnrel \eeact$.
\begin{table}[t!]\centering
$\begin{array}{@{}c@{\qquad}c@{}}
(\eens, \eecnds \cln \eeactsym, \eens') \dclnrel \eeactsym
\quad\text{if $\eens \eectrans{\eecnds \cln \eeactsym} \eens'$}
&
(\eens, \varphi \cln \eempty, \eens) \dclnrel \eecnd{?}
\\\\[-1ex]
\structrule{(\eens, \ewits, \eens') \dclnrel \eeact_{\ell}}{(\eens, \ewits, \eens') \dclnrel \eeact_1 + \eeact_2}
\quad\text{for }\ell \in \{ 1, 2 \}
&
\structrule{(\eens, \ewits_1, \eens') \dclnrel \eeact_1\quad
  (\eens', \ewits_2, \eens'') \dclnrel \eeact_2}{(\eens, \ewits_1 \cdot \ewits_2,  \eens'') \dclnrel \eeact_1; \eeact_2}
\\\\[-1ex]
(\eens, \truefrm \cln \eempty, \eens) \dclnrel \eeact^*
&
\structrule{(\eens, \ewits, \eens') \dclnrel \eeact\quad
  (\eens', \ewits', \eens'') \dclnrel \eeact^*}{(\eens, \ewits \cdot \ewits', \eens'') \dclnrel \eeact^*}
\end{array}$
\vspace*{1.5ex}
\caption{Rules for witnessing compound ensemble actions}
\label{tab:ens-actions-rules}
\vspace*{-1.5ex}
\end{table}

\section{Action Models for Agent Actions}\label{sec:action-models}
 
Going to a concrete interpretation of ensembles a crucial step is to assign
meaning to the agent actions occurring in an epistemic ensemble signature. A
direct way would be to fix a domain for modelling states and to associate to
each agent action $\eeactsym$ a relation (or a proper function) which models the
effect of $\eeactsym$.  This has been done, for instance,
in~\cite{fagin-et-al:2003} who associate to (joint) agent actions a so-called
``global state transformer''.  In our approach we pursue a different approach
and use the notion of ``action model'' introduced
in~\cite{baltag-moss-solecki:tark:1998} to provide meaning for agent actions.
The advantage is that action models still have a syntactic flavour and thus are
still independent of the particular denotations used for epistemic states. Most
approaches use action models in the context of Kripke structures as mathematical
objects for epistemic states on which updates caused by actions are defined.
But, in principle, action models allow also other domains for interpretation.
For instance, in~\cite{Hennicker_Knapp_Wirsing:LPAR:2024} we have provided an
interpretation using symbolic states.  Therefore, our idea is here to use action
models as an intermediate step to assign an ``abstract'' interpretation to agent
actions.  Then we will consider two concrete frameworks (a Kripke style and a
symbolic approach) where updates caused by action model applications are
uniquely determined.
 
An \emph{action model} $\eastr = (\eapnts, \eaacc, \eapre)$ over an epistemic
signature $\ESig = (\Prop, \Ag)$ consists of a set $\eapnts$ of \emph{events},
an $\Ag$-family $\eaacc = (\eaacc_{\ag} \subseteq \eapnts \times \eapnts)_{\ag
  \in \Ag}$ of \emph{action accessibility relations} $\eaacc_{\ag}$, and an
\emph{action precondition} function $\eapre : \eapnts \to \EFrm$.  We assume
that the accessibility relations $\eaacc_{\ag}$ are equivalences.  For any agent
$\ag \in \Ag$, $(\eapnt, \eapnt') \in \eaacc_{\ag}$ models that $\ag$ cannot
distinguish between occurrences of events $\eapnt$ and $\eapnt'$.  For any event
$\eapnt \in \eapnts$, the epistemic formula $\eapre(\eapnt)$ determines a
condition under which $\eapnt$ can happen.
An \emph{epistemic action} $\eact = (\eastr, \eapnt)$ is a pointed action model
which selects an actual event $\eapnt \in \eapnts$.
We set $\eaacc(\eact)_{\ag} = \{ \eapnt' \in \eapnts \mid (\eapnt, \eapnt') \in
\eaacc_{\ag} \}$ for $\ag \in \Ag$, and write $\eapre(\eact)$ for
$\eapre(\eapnt)$ and $\eact \cdot \eapnt'$ for $(\eastr, \eapnt')$ when $\eapnt'
\in \eapnts$.  The class of epistemic actions over $\ESig$ is denoted by
$\EAct$.  The set of possible agents for an epistemic action $\eact \in \EAct$
is defined by $\ags(\eact) = \ags(\eapre(\eact))$.  The idea is that an action
$\eeactsym \in \EEActSym.\ag$ of an agent $a$ should only be interpreted by an
epistemic action whose precondition is an $a$-formula.

\begin{example}\label{ex:anns}
A \emph{group announcement} of a formula $\efrm \in \EFrm$ to a group $\Ag_*
\subseteq \Ag$ of agents is modelled by the epistemic action
$(\eastr_{\grpann}(\Ag_*, \efrm), \eapnt[k])$ graphically represented as
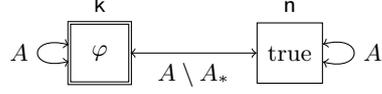
\begin{equation*}
\begin{tikzpicture}[%
  auto, scale=1, every loop/.style={-latex'},%
  font={\fontsize{9pt}{9pt}\selectfont},%
  every label/.append style={outer sep=0pt, inner sep=2pt, label distance=4pt},%
]
\tikzstyle{point}=[rectangle,draw,inner sep=4pt,minimum size=.8cm]
\tikzstyle{access}=[draw,-latex']
\tikzstyle{annotation}=[]
\node[point, double, label={[anchor=base]above:$\eapnt[k]$}] (N) {$\efrm$};
\node[point, label={[anchor=base]above:$\eapnt[n]$}] (P) at ($ (N) + (2.5, 0) $) {$\truefrm$};
\path[access, <->]
(N) edge[loop left, distance=16pt] node[left] {$\Ag$} (N)
(N) edge node[below] {$A \setminus \Ag_*$} (P)
(P) edge[loop right, distance=16pt] node[right] {$\Ag$} (P);
\end{tikzpicture}
\end{equation*}
The action model $\eastr_{\grpann}(\Ag_*, \efrm)$ has two events $\eapnt[k]$ and
$\eapnt[n]$. Event $\eapnt[k]$ represents the announcement of $\efrm$ which
should only happen if $\varphi$ holds and therefore $\eapre_{\grpann,
  \efrm}(\eapnt[k]) = \varphi$; only agents in the group $\Ag_*$ can recognise
this event.  All other agents consider it possible that nothing happened which
is represented by $\eapnt[n]$ having no proper precondition, \ie,
$\eapre_{\grpann, \efrm}(\eapnt[n]) = \truefrm$.  The possible agents of
$(\eastr_{\grpann}(\Ag_*, \efrm), \eapnt[k])$ are given by $\ags(\efrm)$, the
possible agents of $(\eastr_{\grpann}(\Ag_*, \efrm), \eapnt[n])$ are
$\ags(\truefrm) = \Ag$.
\end{example}

\paragraph{Epistemic choice actions.}
We also consider non"=deterministic epistemic actions, similarly
to~\cite{van-ditmarsch-van-der-hoek-kooi:2008}.  They model alternatives which
are not under the control of an agent but are selected by the environment.
Formally, an \emph{epistemic choice action} over $\Sigma$ is a finite,
non"=empty set $\ecact \subseteq \EAct$ of epistemic actions.  The class of
epistemic choice actions over $\Sigma$ is denoted by $\ECAct$.  The possible
agents of an $\ecact \in \ECAct$ are $\ags(\ecact) = \bigcap_{\eact \in \ecact}
\ags(\eact)$.

\begin{example}\label{ex:ecact}
For agents $a, a' \in \Ag$, the epistemic choice action
\begin{equation*}
\sndlos{\ag}{\ag'}{\efrm_{\ag}} = \{ (\eastr_{\grpann}(\{ \ag' \}, \efrm_{\ag}), \eapnt[k]), (\eastr_{\grpann}(\{ \ag' \},
\efrm_{\ag}), \eapnt[n]) \}
\end{equation*}
models a \emph{lossy sending} of the information $\efrm_{\ag} \in \EFrm|\ag$
from $\ag$ to agent $\ag'$, in the sense that afterwards $\ag'$ may know
$\efrm_{\ag}$ if $\ag'$ is aware of the announcement (modelled by
$\eastr_{\grpann}(\{ \ag' \}, \efrm_{\ag}), \eapnt[k])$), but need not if the
announcement is lost (modelled by $\eastr_{\grpann}(\{ \ag' \}, \efrm_{\ag}),
\eapnt[n])$). In any case, $\ag$ cannot recognise whether the announcement was
successful.  Similarly, a \emph{reliable sending} is defined by the (singleton)
choice action $\sndrel{\ag}{\ag'}{\efrm_{\ag}} = \{ (\eastr_{\grpann}(\{ \ag,
\ag' \}, \efrm_{\ag}), \eapnt[k]) \}$ which is modelled by a group announcement
of $\efrm_{\ag}$ such that both agents recognise that the message did work.  It
holds that $\ags(\sndlos{\ag}{\ag'}{\efrm_{\ag}}) =
\ags(\sndrel{\ag}{\ag'}{\efrm_{\ag}}) = \{\ag\}.$
\end{example}

Let us now come back to our epistemic ensemble signature $\EESig = (\ESig,
(\EEActSym.\ag)_{\ag \in \Ag})$ with $\ESig = (\Prop, \Ag)$.  An \emph{epistemic
  action interpretation} for the agent actions in $\bigcup\EEActSym$ is a
function $\eecact: \bigcup\EEActSym \to \ECAct$ such that for all $\ag \in \Ag$
and $\eeactsym \in \EEActSym.\ag$ it holds $\ag \in \ags(\eecact(\eeactsym))$.

\begin{example}\label{ex:witzel-ens-act}
For the agent actions in~\cref{ex:witzel-ens-2} we use the epistemic action
interpretation $\eecact_2$ with $\eecact_2(\mathit{tell}12(\prop[x_1])) =
\sndlos{\ag[1]}{\ag[2]}{\K_1{\prop[x_1]}}$,
$\eecact_2(\mathit{tell}12(\neg\prop[x_1])) =
\sndlos{\ag[1]}{\ag[2]}{\K_1{\neg\prop[x_1]}}\}$ and $\eecact_2(\mathit{stop}) =
(\eastr_{\grpann}(\{\ag[1],\ag[2]\}, \truefrm), \eapnt[k])$ is the group
announcement of $\truefrm$.  Moreover, $\eecact_2(\mathit{ack}21(x_1)) =
\sndrel{\ag[2]}{\ag[1]}{\K_2{x_1}}$ is the reliable sending from agent $\ag[2]$
to agent $\ag[1]$ that agent $\ag[2]$ knows the value of $\prop[x_1]$.  (Recall
that $\K_2{x_1}$ abbreviates $\K_2{\prop[x_1]} \lor \K_2{\neg\prop[x_1]}$.)
\end{example}

\section{Epistemic Ensembles in a Semantic Environment}\label{sec:eens-semantic}

To execute an epistemic ensemble we need an environment where knowledge formulæ,
in particular process guards and change of knowledge caused by agent
actions, are interpreted. This section is based on the traditional possible
worlds model of epistemic logic; see, \eg,~\cite{fagin-et-al:2003}.

\paragraph{Epistemic states.}
An \emph{epistemic structure} $\estr = (\ewrlds, \eacc, \elab)$ (also called
Kripke
model~\cite{baltag-renne:stanford:2016,van-ditmarsch-van-der-hoek-kooi:2008} or
Kripke structure~\cite{fagin-et-al:2003}) over the epistemic signature $\ESig =
(\Prop, \Ag)$ is given by a set $\ewrlds$ of \emph{worlds}, an $\Ag$-family
$\eacc = (\eacc_{\ag} \subseteq \ewrlds \times \ewrlds)_{\ag \in \Ag}$ of
epistemic \emph{accessibility relations}, and a \emph{labelling} $\elab :
\ewrlds \to \powerset \Prop$ which determines for each world $\ewrld \in
\ewrlds$ the set of atomic propositions which hold in $\ewrld$.  We assume that
the accessibility relations are equivalences.  For any agent $\ag$, $(\ewrld,
\ewrld') \in \eacc_{\ag}$ models that agent $\ag$ cannot distinguish the two
worlds $\ewrld$ and $\ewrld'$.  An \emph{epistemic state} $\est = (\estr,
\ewrld)$ selects a world $\ewrld \in \ewrlds$ considered as the \emph{actual}
world. Thus epistemic states are pointed Kripke structures.  The class of
epistemic states over $\ESig$ is denoted by $\ESt$.

The \emph{satisfaction} of an epistemic formula $\efrm \in \EFrm$
by an epistemic structure $\estr = (\ewrlds, \eacc, \elab) \in \ESt$ at a world $\ewrld \in \ewrlds$,
written $\estr, \ewrld \emodels \efrm$, is inductively defined by:
\begin{gather*}
  \estr, \ewrld \emodels \prop \iff \prop \in \elab(\ewrld)
\\
  \estr, \ewrld \emodels \truefrm
\\
  \estr, \ewrld \emodels \neg\efrm \iff \text{not $\estr, \ewrld \emodels \efrm$}
\\
  \estr, \ewrld \emodels \efrm_1 \land \efrm_2 \iff \text{$\estr, \ewrld \emodels  \efrm_1$ and $\estr, \ewrld \emodels \efrm_2$}
\\
  \estr, \ewrld \emodels \K{\efrm} \iff \text{$\estr, \ewrld' \emodels \efrm$ for all $\ewrld' \in \ewrlds$ with $(\ewrld, \ewrld') \in \eacc_{\ag}$}
\end{gather*}
Hence, an agent $\ag$ knows $\efrm$ at point $\ewrld$ if $\efrm$ holds in all
worlds $\ewrld'$ which $\ag$ cannot distinguish from $\ewrld$.  For an epistemic
state $\est = (\estr,\ewrld) \in \ESt$ and for $\efrm \in \EFrm$, we define
$\est \models \efrm$ by $\estr, \ewrld \models \efrm$ and for $\efrms \subseteq
\EFrm$ we define $\est \models \efrms$ by $\est \emodels \efrm$ for all $\efrm
\in \efrms$.  A formula $\efrm \in \EFrm$ is a \emph{logical consequence} of
$\efrms \subseteq \EFrm$, written $\efrms \emodels \efrm$, if $\est \emodels
\efrm$ for all $\est$ with $\est \emodels \efrms$.  We write $\emodels \efrm$
for $\emptyset \emodels \efrm$ , \ie, $\efrm$ is a \emph{tautology}.

\begin{example}\label{ex:witzel-est}
The following diagram represents graphically an epistemic state $\est_0 =
(\estr_0, \ewrld_0)$ in which $\prop[x_1]$ is true and agent $\ag[1]$ knows
this, but agent $\ag[2]$ does not.  Indeed, agent $\ag[2]$ cannot distinguish
between the actual world $\ewrld_0$ and the possible world $\ewrld_1$.  The
self-loops represent reflexivity of the accessibility relations.  Note that
$\est_0 \emodels_{2} \K_1{\prop[x_1]}$, $\est_0 \emodels_{2}
\neg\K_2{\prop[x_1]}$, and $\est_0 \emodels_{2} \K_1{\neg\K_2{\prop[x_1]}}$.
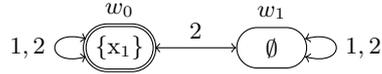
\begin{equation*}
\begin{tikzpicture}[%
  auto, scale=1, every loop/.style={-latex'},%
  font={\fontsize{9pt}{9pt}\selectfont},%
  every label/.append style={outer sep=0pt, inner sep=2pt, label distance=4pt},%
]
\tikzstyle{state}=[rounded rectangle,draw,inner sep=4pt,minimum width=1.2cm]
\tikzstyle{access}=[draw,-latex']
\tikzstyle{annotation}=[]
\node[state,double, label={[anchor=base]above:$\ewrld_0$}] (P) {$\{\prop[x_1]\}$};
\node[state, label={[anchor=base]above:$\ewrld_1$}] (N) at ($ (P) + (2.0, 0) $) {$\emptyset$};
\path[access, <->]
(P) edge[loop left, distance=16pt] node[left] {$\ag[1],\ag[2]$} (P)
(P) edge node[above] {$\ag[2]$} (N)
(N) edge[loop right, distance=16pt] node[right] {$\ag[1],\ag[2]$} (N);
\end{tikzpicture}
\end{equation*}
A state $\est_0'$ with $\est_0' \emodels \K_1{\neg\prop[x_1]}$ and $\est_0'
\emodels \K_1{\neg\K_2{\neg\prop[x_1]}}$ is obtained by reversing the labelling
of $\estr_0$.
\end{example}

\paragraph{Epistemic updates.}
The \emph{product update} $(\ewrlds, \eacc, \elab) \eupd (\eapnts, \eaacc,
\eapre)$ of an epistemic structure $\estr = (\ewrlds, \eacc, \elab)$ and an
action model $\eastr = (\eapnts, \eaacc, \eapre)$ over $\ESig = (\Prop, \Ag)$
yields the epistemic structure $(\ewrlds', \eacc', \elab')$ with
\begin{gather*}
  \ewrlds' = \{ (\ewrld, \eapnt) \in \ewrlds \times \eapnts \mid \estr, \ewrld \emodels \eapre(\eapnt) \}
\ ,\\
  \eacc_{\ag}' = \{ ((\ewrld, \eapnt), (\ewrld', \eapnt')) \mid (\ewrld, \ewrld') \in \eacc_a,\ (\eapnt, \eapnt') \in F_{\ag} \} \text{ for all $\ag \in \Ag$, and}\\
  \elab'(\ewrld, \eapnt) = \elab(\ewrld) \text{ for all $(\ewrld, \eapnt) \in \ewrlds'$.}
\end{gather*}

Let $\est = (\estr, \ewrld) \in \ESt$ be an epistemic state and $\eact =
(\eastr, \eapnt) \in \EAct$ be an epistemic action.  If $\est \emodels
\eapre(\eapnt)$ then the \emph{product update} $\est \eupd \eact$ of $\est$ and
$\eact$ is defined and given by the epistemic state $(\estr \eupd \eastr,
(\ewrld, \eapnt)) \in \ESt$.

\begin{example}\label{ex:witzel-upd}
Applying $(\eastr_{\grpann}(\{\ag[2]\},\K_1{\prop[x_1]}),\eapnt[k])$ to the
epistemic state $(\estr_0, \ewrld_0)$ in~\cref{ex:witzel-est}, we obtain the
epistemic state $(\estr_1, (\ewrld_0,\eapnt[k]))$ shown, without reflexive
accessibility edges, below.  The world $(\ewrld_1, \eapnt[k])$ does not appear
since $(\estr_0, \ewrld_1) \not\emodels_{2} \K_1{\prop[x_1]}$ which is the
precondition of $\eapnt[k]$.
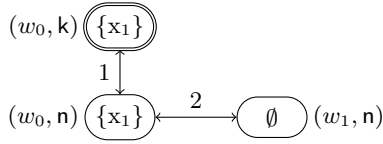
\begin{equation*}
\begin{tikzpicture}[%
  auto, scale=1, every loop/.style={-latex'},%
  font={\fontsize{9pt}{9pt}\selectfont},%
  every label/.append style={outer sep=0pt, inner sep=2pt, label distance=4pt},%
]
\tikzstyle{state}=[rounded rectangle,draw,inner sep=4pt,minimum width=1.2cm]
\tikzstyle{access}=[draw,-latex']
\tikzstyle{annotation}=[]
\node[state, label={[anchor=east, label distance=0pt]left:$(\ewrld_0, \eapnt[n])$}] (Pn) {$\{\prop[x_1]\}$};
\node[state, label={[anchor=west, label distance=0pt]right:$(\ewrld_1, \eapnt[n])$}] (Nn) at ($ (Pn) + (2.0, 0) $) {$\emptyset$};
\node[state, double, label={[anchor=east, label distance=0pt]left:$(\ewrld_0, \eapnt[k])$}] (Pp) at ($ (Pn) + (0, +1.2) $) {$\{\prop[x_1]\}$};
\path[access, <->]
(Pn) edge node[above] {$\ag[2]$} (Nn)
(Pn) edge node[left] {$\ag[1]$} (Pp)
;
\end{tikzpicture}
\end{equation*}
Note that $(\estr_1, (\ewrld_0, \eapnt[k])) \emodels_{2} \K_2{\K_1{\prop[x_1]}}$
but $(\estr_1, (\ewrld_0,\eapnt[k])) \emodels_{2}
\neg\K_1{\K_2{\K_1{\prop[x_1]}}}$, \ie, $\ag[2]$ knows that $\ag[1]$ knows that
$\prop[x_1]$ holds, but $\ag[1]$ does not know that $\ag[2]$ knows this.  If we
apply the epistemic action
$(\eastr_{\grpann}(\{\ag[2]\},\K_1{\prop[x_1]}),\eapnt[n])$ to $(\estr_0,
\ewrld_0)$, we obtain the epistemic state $(\estr_1, (\ewrld_0, \eapnt[n]))$.
Note that $(\estr_1, (\ewrld_0, \eapnt[n])) \emodels_{2}
\neg\K_2{\K_1{\prop[x_1]}}$.
\end{example}

\paragraph{Semantic environments.}
In contrast
to~\cite{hennicker-knapp-wirsing:isola:2022,Hennicker_Knapp_Wirsing:LPAR:2024}
we consider here not only single epistemic states $\est \in \ESt$ as semantic
environments but, more generally, non"=empty classes $\ests \subseteq \ESt$ of
epistemic states; the class of all non"=empty classes of epistemic states is
denoted by $\ESts$.  Such classes model abstractions where different epistemic
states are considered to be possible in a current environment.  For each $\ests
\subseteq \ESt$ and $\efrm \in \EFrm$ define $\ests \emodels \efrm$ as $\est
\emodels \efrm$ for all $\est \in \ests$, and for each $\eact \in \EAct$ define
the update $\ests \eupd \eact = \{ \est \eupd \eact \mid \est \in \ests\text{,}\
\est \emodels \eapre(\eact) \}$.  The \emph{semantics} of an epistemic choice
action $\ecact \in \ECAct$ is given by the relation
\begin{equation*}\textstyle
\esem{\ecact} = \{ (\ests, \ests \eupd \eact) \in \ESts \times \ESts \mid \eact \in \ecact\text{, }\ests \emodels \eapre(\eact) \}
\ \text{.}
\end{equation*}
We require $\ests \neq \emptyset$ because otherwise all formulæ would hold in
$\ests$.  Note that the semantics is well-defined, \ie, $\ests \eupd \eact \neq
\emptyset$, since $\emptyset \neq \ests$ and $\ests \emodels \eapre(\eact)$
hold.  Let us remark that updates on a single epistemic state as considered
in~\cite{hennicker-knapp-wirsing:isola:2022,Hennicker_Knapp_Wirsing:LPAR:2024}
and its generalisation to classes of epistemic states considered here are
tightly related: For each $\est \in \ESt$,
\begin{equation*}\textstyle
\bigcup_{\eact \in \ecact} \{ (\{ \est \}, \{ \est \eupd \eact \}) \mid \est \emodels \eact \} \subseteq \esem{\ecact}
\ \text{.}
\end{equation*}

More generally, we obtain

\begin{restatable}{lemma}{LemECActEStsESt}\label{lem:ecact-ests-est}
Let $\ests \in \ESts$ and $\ecact \in \ECAct$.
\begin{enumerate}
  \item\label{it:lem:ecact-ests-est:zig} If $(\{ \est \}, \ests^{\est}) \in
\esem{\ecact}$ for some $\est \in \ests$, then there are $\est' \in \ESt$ with
$\ests^{\est} = \{ \est' \}$ and $\ests' \in \ESts$ with $\est' \in \ests'$ such
that $(\ests, \ests') \in \esem{\ecact}$.

  \item\label{it:lem:ecact-ests-est:zag} If $(\ests, \ests') \in \esem{\ecact}$, then for each $\est' \in \ests'$ there
is some $\est \in \ests$ with $(\{ \est \}, \{ \est' \}) \in \esem{\ecact}$.
\end{enumerate}
\end{restatable}

\paragraph{Ensemble configurations.}
We are now ready to define the execution of ensembles in semantic class
environments.  Given the epistemic ensemble signature $\EESig = (\ESig,
(\EEActSym.\ag)_{\ag \in \Ag})$, an epistemic \emph{ensemble configuration} over
$\EESig$ is a pair $(\eens, \ests)$ of an ensemble over $\EESig$ and a
(non"=empty) class of epistemic states $\ests \in \ESts$.  The
\emph{ensemble semantics} over $\EESig$ w.r.t.\ an epistemic action
interpretation $\eecact: \bigcup\EEActSym \to \ECAct$ is the ternary relation
$\eetrans{}$ between configurations, agent actions and (successor)
configurations defined by interpreting the syntactic operational ensemble
semantics, given by rule \cref{eq:sem-ens} in~\cref{sec:eensembles}, in the
environment of classes of epistemic states:
\begin{gather*}
\eens, \ests \eetrans{\eeactsym} \eens', \ests'
\quad\text{if }\eens \eectrans{\eecnds \cln \eeactsym} \eens'
\text{, }\ests \emodels \eecnds
\text{, and }(\ests, \ests') \in \esem{\eecact(\eeactsym)}
\end{gather*}

\begin{example}\label{ex:witzel-ests}
Consider the bit transmission protocol of \cref{ex:witzel-ens} with its agent
actions interpreted as in \cref{ex:witzel-ens-act}.  As initial classes of
epistemic states consider on the one hand $\ests_{0, \prop[x_1]}$ with all its
states satisfying $\K_1{\prop[x_1]}$ (as illustrated by $\est_0$ in
\cref{ex:witzel-est}), and $\ests_{0, \neg\prop[x_1]}$ with all its states
satisfying $\K_1{\neg\prop[x_1]}$ (as illustrated by $\est_0'$ in
\cref{ex:witzel-est}).  In both cases, the value of $\prop[x_2]$ is arbitary, as
is the knowledge of the agents $\ag[1]$ and $\ag[2]$ about it.  The ensemble
works uniformly on $\ests_{0, \prop[x_1]}$ and on $\ests_{0, \neg\prop[x_1]}$.
Taking an initial class of epistemic states where some states satisfy
$\K_1{\prop[x_1]}$ and others $\K_1{\neg\prop[x_1]}$ no transition can be taken.
When starting the ensemble in a class of epistemic states all satisfying
$\K_1{\K_2{x_1}}$, then $\mathit{ack}12(x_1)$ and $\mathit{stop}$ can be
executed.
\end{example}

\paragraph{Satisfaction of epistemic ensemble formulæ.}
Finally we are now also able to define when an epistemic ensemble formula (cf.\
\cref{sec:elogic}) is satisfied by an ensemble configuration.  As a preparation
we define the semantics of guard-action sequences $\ewits = (\eecnds_1 \cln
\eeactsym_1^{\eempty}) \ldots (\eecnds_k \cln \eeactsym_k^{\eempty})$ relating
classes of epistemic states inductively by
\begin{gather*}
\eesem{\eecnds \cln \eempty} = \{ (\ests, \ests) \in \ESts \times \ESts \mid \ests \emodels \eecnds \}
\ \text{,}
\\
\eesem{\eecnds \cln \eeactsym} = \{ (\ests, \ests') \in \esem{\eecact(\eeactsym)} \mid \ests \emodels \eecnds \}
\ \text{,}
\\
\eesem{\ewits_1 \cdot \ewits_2} = \eesem{\ewits_1}; \eesem{\ewits_2} 
\quad\text{(relational composition)}
\end{gather*}
The semantics of a compound ensemble
action $\eeact \in \EEAct$ relating ensemble configurations relies on the definition of witnesses
in~\cref{sec:elogic} and is given by the relation
\begin{equation*}
\eesem{\eeact} = \{ ((\eens, \ests), (\eens', \ests')) \mid (\eens, \ewits, \eens') \dclnrel \eeact,\ (\ests, \ests') \in \eesem{\ewits} \}
\ \text{.}
\end{equation*}

\begin{restatable}{lemma}{EEActEStsESt}\label{lem:eeact-ests-est}
Let $(\eens, \ests)$ be an ensemble configuration and $\eeact \in \EEAct$.
\begin{enumerate}
  \item\label{it:lem:eeact-ests-est:zig} If $((\eens, \{ \est \}), (\eens',
\ests^{\est})) \in \eesem{\eeact}$ for some $\est \in \ests$, then there are
$\est' \in \ESt$ with $\ests^{\est} = \{ \est' \}$ and $\ests' \in \ESts$ with
$\est' \in \ests'$ such that $((\eens, \ests), (\eens', \ests')) \in
\eesem{\eeact}$.

  \item\label{it:lem:eeact-ests-est:zag} If $((\eens, \ests), (\eens, \ests')) \in \eesem{\eeact}$, then for each $\est' \in \ests'$ there
is some $\est \in \ests$ with $((\eens, \{ \est \}), (\eens', \{ \est' \})) \in \eesem{\eeact}$.
\end{enumerate}
\end{restatable}

The satisfaction of an epistemic ensemble formula $\eefrm \in \EEFrm$ by an
ensemble configuration $(\eens, \ests)$
w.r.t.\ $\eecact: \bigcup\EEActSym \to \ECAct$ is inductively defined along
the form of $\eefrm$:
\begin{gather*}
(\eens, \ests) \eemodels \efrm \iff \ests \emodels \efrm
\\
(\eens, \ests) \eemodels \truefrm
\\
(\eens, \ests) \eemodels \neg\eefrm
\iff \text{not } (\eens, \ests) \eemodels \eefrm
\\
(\eens, \ests) \eemodels \eefrm_1 \land \eefrm_2
\iff (\eens, \ests) \eemodels \eefrm_1 \text{ and } (\eens, \ests) \eemodels \eefrm_2
\\
(\eens, \ests) \eemodels \dlbox{\eeact}{\eefrm}
\iff \stacked{(\eens', \ests') \eemodels \eefrm\\ \text{for all }
(\eens', \ests') \text{ such that } ((\eens, \ests), (\eens', \ests')) \in \eesem{\eeact}}
\end{gather*}

\begin{example}
The ensemble configuration $(\mathit{Sys}, \ests_{0, \prop[x_1]})$ for the bit
transmission ensemble $\mathit{Sys}$ from \cref{ex:witzel-ens} and the class of
epistemic states $\ests_{0, \prop[x_1]}$ from \cref{ex:witzel-ests} satisfies
both dynamic ensemble formulæ of~\cref{ex:elogic}; the same is true for
$(\mathit{Sys}, \ests_{0, \prop[x_1]})$.
\end{example}

\begin{restatable}{proposition}{PropEEFrmEStsEst}\label{prop:eefrm-ests-est}
Let $(\eens, \ests)$ be an ensemble configuration.  For all $\eefrm \in \EEFrm$ it holds that $(\eens, \ests) \eemodels \eefrm$ if, and only if,
$(\eens, \{ \est \}) \eemodels \eefrm$ for all $\est \in \ests$.
\end{restatable}

\section{Epistemic Ensembles in a Symbolic Environment}\label{sec:eens-symbolic}

We are now going to execute ensembles in a symbolic environment which allows for
a more compact representation of epistemic states represented by sets of
epistemic formulæ, \ie, knowledge bases.  For doing this we extend the approach
of~\cite{Hennicker_Knapp_Wirsing:LPAR:2024} to deal with ensembles.  A
\emph{symbolic epistemic signature} $\SESig$ extends the epistemic signature
$\ESig = (\Prop,\Ag)$ by a finite set of epistemic formulæ $\Fcs \subseteq
\EFrm$ which are in the \emph{focus} of evaluation.  Focusing to a finite set
allows for effective computations and decisions.

\paragraph{Symbolic epistemic states.}
A \emph{symbolic epistemic state} over $\SESig$ is a subset $\sest \subseteq
\Fcs$ which is \emph{$\truefrm$"=closed}, \ie, if $\truefrm \in \Fcs$, then
$\truefrm \in \sest$.  (If $\truefrm \in \Fcs$, but $\truefrm \notin \sest$,
this would mean that we consider $\truefrm$ not to hold.).  The set of symbolic
epistemic states over $\SESig$ is denoted by $\SESt$.

The \emph{Boolean closure} $\bcl(\Fcs)$ of $\Fcs$ consists of the epistemic
formulæ $\befrm$ defined by
\begin{equation*}
  \befrm
{\;\cln\cln=\;}
  \varphi \;\mid\;
  \truefrm \;\mid\;
  \neg\befrm \;\mid\;
  \befrm_1 \land \befrm_2 \qquad\text{where $\efrm \in \Fcs$.}
\end{equation*}
The \emph{symbolic satisfaction relation} $\sest \semodels \befrm$
between symbolic states $\sest \in \SESt$ and formulæ $\befrm \in \bcl(\Fcs)$ is defined as:
\begin{align*}
\text{if $\efrm \in \Fcs$: } &
\sest \semodels \efrm
\iff \efrm \in \sest
\\
& \sest \semodels \truefrm
\\
\text{if $\neg\befrm \notin \Fcs$: } &
\sest \semodels \neg\befrm
\iff \text{not } \sest \semodels \befrm
\\
\text{if $\befrm_1 \land \befrm_2 \notin \Fcs$: } &
\sest \semodels \befrm_1 \land \befrm_2
\iff \sest \semodels \befrm_1 \text{ and } \sest \semodels \befrm_2
\end{align*}

\paragraph{Symbolic epistemic updates.}
Let $\eact \in \EAct$ be an epistemic action and $\efrm \in \EFrm$.  A formula
$\erepr \in \EFrm$ is a \emph{weakest liberal precondition} of $\eact$ for
  $\efrm$ if the following holds for all $\est \in \ESt$:
\begin{equation}\label{eq:wlp}\tag{wlp}
\est \emodels \erepr \iff \big(\est \emodels \eapre(\eact) \text{ implies }
\est \eupd \eact \emodels \efrm\big)
\end{equation}
The set of the weakest liberal precondition formulæ of $\eact$ for $\efrm$ is
denoted by $\Wlp{\eact}{\efrm}$.  Obviously, if $\erepr, \erepr' \in
\Wlp{\eact}{\efrm}$, then $\emodels \erepr \lequiv \erepr'$. There is indeed,
for any $\eact \in \EAct$ and any $\efrm \in \EFrm$, a formula
$\wlp{\eact}{\efrm} \in \Wlp{\eact}{\efrm}$ that can be recursively computed by
the function $\wlpop: \EAct \times \EFrm \to \EFrm$ defined in accordance with
the reduction rules originally stated in the context of dynamic epistemic logic
(DEL) in~\cite[pp.~162sqq.]{van-ditmarsch-van-der-hoek-kooi:2008}
and~\cite[p.~37]{baltag-renne:stanford:2016}:
\begin{gather*}
\wlp{\eact}{\prop} = \eapre(\eact) \limp \prop
\\
\wlp{\eact}{\truefrm} = \truefrm
\\
\wlp{\eact}{\neg\efrm} = \eapre(\eact) \limp \neg\wlp{\eact}{\efrm}
\\
\wlp{\eact}{\efrm_1 \land \efrm_2} = \wlp{\eact}{\efrm_1} \land  \wlp{\eact}{\efrm_2}
\\\textstyle
\wlp{\eact}{\K{\efrm}} = \eapre(\eact) \limp \bigwedge_{\eapnt \in \eaacc(\eact)_{\ag}}\K{\wlp{\eact \cdot\eapnt}{\efrm}}
\end{gather*}

The \emph{symbolic epistemic update} $\sest \seupd \eact$ of a symbolic
epistemic state $\sest \in \SESt$ by an epistemic action $\eact \in
\SEAct$ is defined as
\begin{gather*}
\sest \seupd \eact = \{ \varphi \in \Fcs \mid \text{ex.\ } \erepr \in \Wlp{\eact}{\efrm} \cap \bcl(\Fcs) \text{ s.\,t.\ } \sest \semodels \erepr \}
\ \text{.}
\end{gather*}

\paragraph{Symbolic environments.}
For an epistemic action $\eact \in \EAct$ define $\eaPre(\eact) = \{ \erepr \in
\EFrm \mid {}\emodels \eapre(\eact) \lequiv \erepr \}$.  We say that $\eact \in
\EAct$ is \emph{$\Fcs$-representable} if (1)~$\eaPre(\eact) \cap \bcl(\Fcs)
\neq \emptyset$ and if (2)~for all $\efrm \in \Fcs$ it holds that
$\Wlp{\eact}{\efrm} \cap \bcl(\Fcs) \neq \emptyset$.  The class of such
epistemic actions is denoted by $\SEAct$.  An epistemic choice action $\ecact
\in \ECAct$ is \emph{$\Fcs$"=representable} if all $\eact \in \ecact$ are
$\Fcs$"=representable; the class of such epistemic choice actions is denoted by
$\SECAct$.

\begin{example}\label{ex:witzel-symb}
For the scenario of \cref{ex:witzel-ens} consider the preliminary focus formulæ
$\Fcs_{2}^{(0)} = \{ \K_1{\prop[x_1]}, \K_1{\neg\prop[x_1]},
\K_2{x_1},\allowbreak \K_1{\K_2{x_1}} \}$ (as before we write $\K{x_1}$ for
$\K{\prop[x_1]} \lor \K{\neg\prop[x_1]}$).  The epistemic actions occuring in
$\sndlos{1}{2}{\K_1{\prop[x_1]}}$, $\sndlos{1}{2}{\K_1{\neg\prop[x_1]}}$, and
$\sndrel{2}{1}{\K_2{x_1}}$ are $\eact_{\efrm}^{\eapnt} = (\eastr_{\grpann}(\{
\ag[2] \}, \K_1{\efrm}), \eapnt)$ for $\efrm \in \{ \prop[x_1], \neg\prop[x_1]
\}$, $\eapnt \in \{ \eapnt[k], \eapnt[n] \}$ and $\eact_2 = (\eastr_{\grpann}(\{
\ag[1], \ag[2] \},\allowbreak \K_2{x_1}),\allowbreak \eapnt[k])$.  All these
actions have a precondition that is expressible over $\Fcs_2^{(0)}$.
\Cref{tab:witzel-repr} shows possible representatives $\erepr$ satisfying
$\emodels_{2} \wlp_{2}{\eact}{\efrm} \lequiv \erepr$.\footnote{Computed with a small Maude tool available at
  \url{https://github.com/AlexanderKnapp/epistemic.git}.}
\begin{table}[t!]\centering
\begin{tabular}{@{}r@{\quad}l@{\quad}l@{\quad}l@{\quad}l@{\quad}l@{}}
\toprule
\diagbox[height=1.4\line, width=1.2cm]{$\efrm$}{$\eact$}
& $\eact_{\prop[x_1]}^{\eapnt[k]}$
& $\eact_{\neg\prop[x_1]}^{\eapnt[k]}$
& $\eact_{\prop[x_1]}^{\eapnt[n]}$
& $\eact_{\neg\prop[x_1]}^{\eapnt[n]}$
& $\eact_2$
\\
\midrule
$\K_1{\prop[x_1]}$
& $\truefrm$
& $\neg\K_1{\neg\prop[x_1]}$
& $\K_1{\prop[x_1]}$
& $\K_1{\prop[x_1]}$
& $\stacked[c]{\K_2{x_1} \limp{}\\\ \ \K_1{\M_2{\prop[x_1]}}}$
\\
$\K_1{\neg\prop[x_1]}$
& $\neg\K_1{\prop[x_1]}$
& $\truefrm$
& $\K_1{\neg\prop[x_1]}$
& $\K_1{\neg\prop[x_1]}$
& $\stacked[c]{\K_2{x_1} \limp{}\\\ \ \K_1{\M_2{\neg\prop[x_1]}}}$
\\[1.8ex]
$\K_2{x_1}$
& $\truefrm$
& $\truefrm$
& $\K_2{x_1}$
& $\K_2{x_1}$
& $\truefrm$
\\[.5ex]
$\K_1{\K_2{x_1}}$
& $\stacked[c]{\K_1{\prop[x_1]} \limp\\\ \ \K_1{\K_2{x_1}}}$
& $\stacked[c]{\K_1{\neg\prop[x_1]} \limp\\\ \ \K_1{\K_2{x_1}}}$
& $\K_1{\K_2{x_1}}$
& $\K_1{\K_2{x_1}}$
& $\truefrm$
\\
$\K_1{\M_2{\prop[x_1]}}$
& $\truefrm$
& $\neg\K_1{\neg\prop[x_1]}$
& $\K_1{\M_2{\prop[x_1]}}$
& $\stacked[c]{\neg\K_1{\neg\prop[x_1]} \land\\\ \ \K_1{\M_2{\prop[x_1]}}}$
& $\stacked[c]{\K_2{x_1} \limp\\\ \ \K_1{\M_2{\prop[x_1]}}}$
\\
$\K_1{\M_2{\neg\prop[x_1]}}$
& $\neg\K_1{\prop[x_1]}$
& $\truefrm$
& $\stacked[c]{\neg\K_1{\prop[x_1]} \land\\\ \ \K_1{\M_2{\neg\prop[x_1]}}}$
& $\K_1{\M_2{\neg\prop[x_1]}}$
& $\stacked[c]{\K_2{x_1} \limp\\\ \ \K_1{\M_2{\neg\prop[x_1]}}}$
\\
\bottomrule
\end{tabular}
\vspace*{2ex}
\caption{Representatives $\erepr$ satisfying $\emodels_2 \wlp{\eact}{\efrm} \lequiv \erepr$}\label{tab:witzel-repr}
\vspace*{-1.5ex}
\end{table}
Indeed, $\eact_2$ is not $\Fcs_2^{(0)}$"=representable, but becomes
$\Fcs_2$"=representable for $\Fcs_2 = \Fcs_2^{(0)} \cup \{
\K_1{\M_2{\prop[x_1]}},\allowbreak \K_1{\M_2{\neg\prop[x_2]}} \}$.
\end{example}

The \emph{symbolic semantics} of a $\Fcs$"=representable epistemic choice action
$\ecact \in \SECAct$ is given by the relation
\bgroup
\mathindent8pt
\begin{gather*}\textstyle
\sesem{\ecact} = \{ (\sest, \sest \seupd \eact) \in \SESt \times \SESt \mid{}
\eact \in \ecact\text{, ex.\ }\erepr \in \eaPre(\eact) \cap \bcl(\Fcs)\text{ s.\,t.\ }\sest \semodels \erepr \}
\ \text{.}
\end{gather*}
\egroup

A \emph{symbolic epistemic ensemble signature} $(\EESig, \Fcs)$ consists of an
epistemic ensemble signature $(\ESig, (\EEActSym.\ag)_{\ag \in \Ag})$ and a set
of focus formulæ $\Fcs$ such that $(\ESig, \Fcs)$ is a symbolic epistemic
signature.  An ensemble $\eens$ over $\EESig$ is an ensemble over $(\EESig,
\Fcs)$ if all guards occurring in $\eens$ are in $\bcl(\Fcs)$.  An epistemic
action interpretation $\eecact : \bigcup\EEActSym \to \ECAct$ is a
\emph{$\Fcs$"=interpretation} if $\eecact(\eeactsym) \in \SECAct$ for all
$\eeactsym \in \bigcup\EEActSym$.  A symbolic epistemic \emph{ensemble
  configuration} over $(\EESig, \Fcs)$ is a pair $(\eens, \sest)$ of an ensemble
over $(\EESig, \Fcs)$ and a symbolic epistemic state $\sest \in \SESt$.  The
\emph{symbolic ensemble semantics} over $(\EESig, \Fcs)$ w.r.t.\ a
$\Fcs$"=interpretation $\eecact: \bigcup\EEActSym \to \SECAct$ is the ternary
relation $\seetrans{}$ between symbolic configurations, agent actions and
(successor) configurations defined by interpreting the operational ensemble
semantics, given by rule \cref{eq:sem-ens} in~\cref{sec:eensembles}, in the
environment of symbolic epistemic states:
\begin{gather*}
\eens, \sest \seetrans{\eeactsym} \eens', \sest'
\quad\text{if }\eens \eectrans{\beecnds \cln \eeactsym} \eens'
\text{, }\sest \semodels \beecnds
\text{, and }(\sest, \sest') \in \sesem{\eecact(\eeactsym)}
\end{gather*}

\paragraph{Satisfaction of epistemic ensemble formulæ.}
The \emph{symbolic semantics} of a guard"=agent action sequence $\ewits =
(\beecnds_1 \cln \eeactsym_1^{\eempty}) \ldots (\beecnds_k \cln
\eeactsym_k^{\eempty})$ with all guards $\beecnds_i$ formulæ in $\bcl(\Fcs)$
\wrt the $\Fcs$"=interpretation $\eecact$ is inductively given by
\begin{gather*}
\seesem{\beecnds \cln \eempty} = \{ (\sest, \sest) \in (\SESt)^2 \mid \sest \semodels \beecnds \}
\ \text{,}
\\
\seesem{\beecnds \cln \eeactsym} = \{ (\sest, \sest') \in \sesem{\eecact(\eeactsym)} \mid \sest \semodels \beecnds \}
\ \text{,}
\\
\seesem{\ewits_1 \cdot \ewits_2} = \seesem{\ewits_1}; \seesem{\ewits_2} 
\quad\text{(relational composition)}
\end{gather*}
The set of compound ensemble actions $\eeact \in \EEAct$ with all tests in
$\eeact$ formulæ in $\bcl(\Fcs)$ is denoted $\SEEAct$.  The \emph{semantics} of
a $\eeact \in \SEEAct$ \wrt $\eecact$ is given by the relation
\begin{equation*}
\seesem{\eeact} = \{ ((\eens, \sest), (\eens', \sest')) \mid (\eens, \ewits, \eens') \dclnrel \eeact,\ (\sest, \sest') \in \seesem{\ewits} \}
\ \text{.}
\end{equation*}

The epistemic ensemble formulæ $\SEEFrm$ over $(\EESig, \Fcs)$ are the epistemic ensemble formulæ over $\EESig$ that only contain sub"=formulæ $\befrm \in \EFrm$ that are in $\bcl(\Fcs)$.   The satisfaction of a $\eefrm \in \SEEFrm$ by a
symbolic epistemic ensemble configuration $(\eens, \sest)$
w.r.t.\ $\Fcs$"=interpretation $\eecact: \bigcup\EEActSym \to \SECAct$
is inductively
defined along the structure of $\eefrm$:
\begin{gather*}
(\eens, \sest) \seemodels \befrm \iff \sest \semodels \befrm
\\
(\eens, \sest) \seemodels \truefrm
\\
(\eens, \sest) \seemodels \neg\eefrm
\iff \text{not } (\eens, \sest) \seemodels \eefrm
\\
(\eens, \sest) \seemodels \eefrm_1 \land \eefrm_2
\iff (\eens, \sest) \seemodels \eefrm_1 \text{ and } (\eens, \sest) \seemodels \eefrm_2
\\
(\eens, \sest) \seemodels \dlbox{\eeact}{\eefrm}
\iff \stacked{(\eens', \sest') \seemodels \eefrm\\ \text{for all }
(\eens', \sest') \text{ such that } ((\eens, \sest), (\eens', \sest')) \in \seesem{\eeact}}
\end{gather*}

\section{Relating Ensembles in Semantic and Symbolic Environments}\label{sec:eens-semantic-symbolic}

We are now interested in relating ensembles which are executed in a semantic and in a symbolic environment.  We first introduce a notion of equivalence between semantic and symbolic states and then lift this equivalence to ensemble configurations.  Our main theorem states that both simulate each other and consequently equivalent configurations satisfy the same dynamic epistemic ensemble logic formulæ.

\paragraph{Epistemic state equivalence.}
We say that an epistemic state class $\ests \in \ESts$ and a symbolic epistemic state $\sest
\in \SESt$ are \emph{$\Fcs$"=equivalent}, written $\ests \seequiv \sest$, if for
all $\est \in \ests$ and all $\efrm \in \Fcs$ it holds that $\est \emodels
\efrm$ if, and only if, $\efrm \in \sest$.

\begin{restatable}{lemma}{LemSEEquivBCL}\label{lem:seequiv-bcl}
Let $\ests \in \ESts$ and $\sest \in \SESt$.  Then $\ests \seequiv \sest$
if, and only if, for all $\est \in \ests$ and all $\befrm \in \bcl(\Fcs)$ it
holds that $\est \emodels \befrm$ iff $\sest \semodels \befrm$.
\end{restatable}

\vspace*{-1.4ex}
\begin{restatable}{corollary}{CorSEEquivBCL}\label{cor:seequiv-bcl}
Let $\ests \in \ESts$ and $\sest \in \SESt$ with $\ests \seequiv \sest$.  Then
it holds for all $\befrm \in \bcl(\Fcs)$ that $\ests \emodels \befrm$ if, and
only if, $\sest \semodels \befrm$.
\end{restatable}

\vspace*{-1.4ex}
\begin{restatable}{lemma}{LemWLPESts}\label{lem:wlp-ests}
Let $\ests \subseteq \ESt$, $\eact \in \EAct$, and $\efrm \in \EFrm$.  Then
$\ests \emodels \wlp{\eact}{\efrm}$ if, and only if, $\ests \eupd \eact \emodels
\efrm$.
\end{restatable}

\vspace*{-1.4ex}
\begin{restatable}{lemma}{LemEUpdSEUpd}\label{lem:eupd-seupd}
Let $\ests \in \ESts$ and $\sest \in \SESt$ with $\ests \seequiv \sest$. Let
$\eact \in \SEAct$.  Then $\ests \eupd \eact \seequiv \sest \seupd \eact$.
\end{restatable}

\vspace*{-1.4ex}
\begin{restatable}{lemma}{LemECActSem}\label{lem:ecact-sem}
Let $\ests \in \ESt$ and $\sest \in \SESt$ with $\ests \seequiv \sest$.  Let
$\ecact \in \SECAct$. Then the following holds:
\begin{enumerate}
  \item\label{it:lem:ecact-sem:zig} If $(\ests, \ests') \in \esem{\ecact}$, then
there is a $\sest' \in \SESt$ such that $(\sest, \sest') \in \sesem{\ecact}$ and
$\ests' \seequiv \sest'$.
  \item\label{it:lem:ecact-sem:zag} If $(\sest, \sest') \in \sesem{\ecact}$,
then there is a $\ests' \in \ESts$ such that $(\ests, \ests') \in \esem{\ecact}$ and
$\ests' \seequiv \sest'$.
\end{enumerate}
\end{restatable}

\paragraph{Ensemble configuration equivalence.}
As a consequence of~\cref{lem:ecact-sem} we can prove that semantic and symbolic
ensemble configurations mutually simulate action execution while preserving
$\Fcs$"=equivalence.

\begin{restatable}{proposition}{PropEEActSymBisim}\label{prop:eeactsym-bisim}
Let $(\eens, \ests)$ be an epistemic ensemble configuration over $\EESig$ and
$(\eens, \sest)$ a symbolic epistemic ensemble configuration over $(\EESig,
\Fcs)$.  Let $\eeactsym \in \bigcup\EEActSym$ be an agent action, and let $\ests
\seequiv \sest$ hold.
\begin{enumerate}
  \item\label{it:prop:eeactsym-bisim:zig} If $(\eens, \ests) \eetrans{\eeactsym}
(\eens', \ests')$, then there is a $\sest' \in \SESt$ such that $(\eens, \sest)
\seetrans{\eeact} (\eens', \sest')$ and $\ests' \seequiv \sest'$.

  \item\label{it:prop:eeactsym-bisim:zag} If $(\eens, \sest)
\seetrans{\eeactsym} (\eens', \sest')$, then there is a $\ests' \in \ESts$
such that $(\eens, \ests) \eetrans{\eeact} (\eens', \ests')$ and $\ests'
\seequiv \sest'$.
\end{enumerate}
\end{restatable}

The $\Fcs$"=equivalence $\est \seequiv \sest$ between semantic and symbolic
epistemic states induces a $(\Fcs, \eecact)$"=equivalence between semantic and
symbolic epistemic ensemble configurations defined as $(\eens, \ests) \seeequiv
(\eens, \sest)$ if $\ests \seequiv \sest$.  The next proposition
lifts~\cref{prop:eeactsym-bisim} to compound epistemic actions:

\begin{restatable}{proposition}{PropEEActSem}\label{prop:eeact-sem}
Let $\eeact \in \SEEAct$ and let $(\eens, \ests) \seeequiv (\eens, \sest)$
hold.
\begin{enumerate}
  \item\label{it:prop:eeact-sem:zig} If $((\eens, \ests), (\eens', \ests')) \in
\eesem{\eeact}$, then there exists a $\sest' \in \SESt$ such that $((\eens,
\sest),\allowbreak (\eens',\sest')) \in \seesem{\eeact}$ and $(\eens', \ests')
\seeequiv (\eens', \sest')$.

  \item\label{it:prop:eeact-sem:zag} If $((\eens, \sest), (\eens',\sest')) \in
\seesem{\eeact}$, then there exists a $\ests' \in \ESts$ such that
$((\eens, \ests),\allowbreak (\eens',\ests')) \in \eesem{\eeact}$ and $(\eens',
\ests') \seeequiv (\eens', \sest')$.
\end{enumerate}
\end{restatable}

Finally, $(\Fcs, \eecact)$"=equivalent ensemble configurations satisfy the same
dynamic ensemble logic formulæ.  Thus symbolic epistemic ensemble configurations
can be considered as correct realisations of semantic epistemic ensemble
configurations.

\begin{restatable}{theorem}{ThmMain}\label{thm:main}
Let $(\eens, \ests) \seeequiv (\eens, \sest)$ hold.  Then for all $\eefrm \in
\SEEFrm$, it holds that $(\eens, \ests) \eemodels \eefrm \iff (\eens, \sest)
\seemodels \eefrm$.
\end{restatable}

\section{Conclusions}\label{sec:conclusions}

Epistemic ensembles are families of interacting knowledge-based
processes. Starting with a syntactic operational semantics, we presented two
complementary mathematical semantics for such ensembles: one in a semantic
environment defined by classes of global epistemic states and the other one in a
symbolic environment defined by a global knowledge base. As main result we
showed that ensembles with $\Fcs$"=equivalent configurations simulate each other
and preserve $\Fcs$"=equivalence.

In this paper we gave only small examples; in the future we want to tackle
larger applications in behavioural ensemble specification (see,
\eg,~\cite{Surmeli20,hennicker-knapp-wirsing:isola:2022} and epistemic planning (see,
\eg,~\cite{DBLP:journals/corr/Bolander17}). We also intend to extend our
approach to ensembles with distributed local states (see,
\eg,~\cite{BolanderA11}) and to LLM-assisted software
development~\cite{BelznerGW23}.

\paragraph{Acknowledgements.}
We would like to thank anonymous reviewers of this paper for valuable comments. 

\bibliographystyle{plain}
\bibliography{bibliography}

\ifarXiv\else%
\end{document}%
\fi

\begin{appendix}

\section{Proofs}

\LemProcAgs*
\begin{proof}
We proceed by induction on the derivation of the conditional transition $\eeproc
\epctrans{\eecnds \cln \eeactsym} \eeproc'$ by the rules in \cref{tab:proc-rules}.

\smallskip\noindent%
For $\eeactsym.\eeproc \epctrans{\truefrm \cln \eeactsym} \eeproc$, we have $\eeags(\eeactsym.\eeproc) = \eeags(\eeactsym) \cap \eeags(\eeproc) \subseteq \ags(\truefrm) \cap \eeags(\eeactsym) \cap \eeags(\eeproc)$.

\smallskip\noindent%
For $\eecnd \supset \eeproc \epctrans{\eecnds \land \eecnd \cln \eeactsym}
\eeproc'$, we have $\eeags(\eeproc) \subseteq \ags(\eecnds) \cap \eeags(\eeactsym) \cap
\ags(\eeproc')$ as the induction hypothesis and thus we obtain $\eeags(\eecnd
\supset \eeproc) = \ags(\eecnd) \cap \eeags(\eeproc) \subseteq \ags(\eecnd) \cap
\ags(\eecnds) \cap \eeags(\eeactsym) \cap \eeags(\eeproc')
= \ags(\eecnds \land \eecnd) \cap \eeags(\eeactsym) \cap \eeags(\eeproc')$.

\smallskip\noindent%
For $\eeproc_1 + \eeproc_2 \epctrans{\eecnds \cln \eeactsym} \eeproc_{\ell}'$ with
$\ell \in \{ 1, 2 \}$, we have $\eeags(\eeproc_{\ell}) \subseteq \ags(\eecnds)
\cap \eeags(\eeactsym) \cap \eeags(\eeproc_{\ell}')$ as the induction hypothesis and
thus we obtain $\eeags(\eeproc_1 + \eeproc_2) = \eeags(\eeproc_1) \cap
\eeags(\eeproc_2) \subseteq \ags(\eecnds) \cap \eeags(\eeactsym) \cap
\eeags(\eeproc_{\ell}')$.

\smallskip\noindent%
For $\mu X \,.\, \eeproc \epctrans{\eecnds \cln \eeactsym} \eeproc'$, we have
$\eeags(\eeproc\{ X \mapsto \mu X \,.\, \eeproc \}) \subseteq \ags(\eecnds) \cap
\eeags(\eeactsym) \cap \eeags(\eeproc')$ as the induction hypothesis and thus
$\eeags(\mu X \,.\, \eeproc) = \eeags(\eeproc) \subseteq \ags(\eecnds) \cap
\eeags(\eeactsym) \cap \eeags(\eeproc')$ using that $\eeags(\eeproc\{ X \mapsto \mu X
\,.\, \eeproc \}) = \eeags(\eeproc)$.
\end{proof}

\LemECActEStsESt*
\begin{proof}
\cref{it:lem:ecact-ests-est:zig} Let $(\{ \est \}, \ests^{\est}) \in
\esem{\ecact}$ hold.  Then there is a $\eact \in \ecact$ with $\est \emodels
\eapre(\eact)$ and $\ests^{\est} = \{ \est \eupd \eact \}$.  Furthermore, $\ests
\emodels \eapre(\eact)$ and thus $(\ests, \ests \eupd \eact) \in \esem{\ecact}$
with $\est \eupd \eact \in \ests \eupd \eact$.

\smallskip\noindent%
\cref{it:lem:ecact-ests-est:zag} Let $(\ests, \ests') \in \esem{\ecact}$ hold
and let $\est' \in \ests'$.  Then there is a $\eact \in \ecact$ with $\ests
\emodels \eapre(\eact)$ and $\ests' = \ests \eupd \eact$ such that there is
$\est \in \ests$ with $\est' = \est \eupd \eact$ and $(\{ \est \}, \{ \est' \})
\in \esem{\ecact}$.
\end{proof}

\EEActEStsESt*
\begin{proof}
The claims follow directly from \cref{lem:ecact-ests-est} and the definition of the semantics of guard-action sequences $\ewits$.
\end{proof}

\PropEEFrmEStsEst*
\begin{proof}
We proceed by induction over the structure of $\eefrm \in \EEFrm$ where $\efrm$,
$\truefrm$, negation, and conjunction are straightforward.  For a
$\dlbox{\eeact}{\eefrm}$, let first $(\eens, \ests) \eemodels
\dlbox{\eeact}{\eefrm}$ hold.  Let $\est \in \ests$ and $((\eens, \{ \est \}),
(\eens', \ests^{\est})) \in \eesem{\eeact}$.  By
\cref{lem:eeact-ests-est}\cref{it:lem:eeact-ests-est:zig} there is a $\est' \in
\ESt$ with $\ests^{\est} = \{ \est' \}$ and some $\ests' \in \ESts$ with
$\est' \in \ests'$ and $((\eens, \ests), (\eens', \ests')) \in \eesem{\eeact}$
and thus $(\eens', \ests') \eemodels \eefrm$.  By the induction hypothesis,
$(\eens', \{ \est' \}) \eemodels \eefrm$ and thus $(\eens, \{ \est \}) \eemodels
\dlbox{\eeact}{\eefrm}$.~--- Let conversely, $(\eens, \{ \est \}) \eemodels
\dlbox{\eeact}{\eefrm}$ hold for all $\est \in \ests$.  Let $((\eens, \ests),
(\eens', \ests')) \in \eesem{\eeact}$.  By
\cref{lem:eeact-ests-est}\cref{it:lem:eeact-ests-est:zag} for each $\est' \in
\ests'$ there is a $\est \in \ests$ with $((\eens, \{ \est \}), (\eens', \{
\est' \})) \in \eesem{\eeact}$ and thus $(\eens, \{ \est' \}) \emodels \eefrm$
for all $\est' \in \ests'$.  By the induction hypothesis, $(\eens', \ests')
\eemodels \eefrm$ and thus $(\eens, \{ \est \}) \eemodels
\dlbox{\eeact}{\eefrm}$.
\end{proof}

\LemSEEquivBCL*
\begin{proof}
``$\Rightarrow$'': Assume $\ests \seequiv \sest$ and let $\est \in \ests$.  The
proof is performed by structural induction on the form of $\befrm$.

\smallskip\noindent%
Case $\befrm = \efrm \in \Fcs$: Then $\est \emodels \efrm$ iff (since $\ests
\seequiv \sest$) $\efrm \in \sest$ iff $\sest \semodels \efrm$.

\smallskip\noindent%
Case $\truefrm$: $\est \emodels \truefrm$ and $\sest \semodels \truefrm$ hold.

\smallskip\noindent%
Case $\neg\befrm$ with $\neg\befrm \notin \Fcs$: $\est \emodels \neg\befrm$ iff
not $\est \emodels \befrm$ iff (by induction hypothesis) not $\sest \semodels
\befrm$ iff (since $\neg\befrm \notin \Fcs$) $\sest \semodels \neg\befrm$.

\smallskip\noindent%
Case $\varphi_1 \land \varphi_2$ with $\varphi_1 \land \varphi_2 \notin \Fcs$:
$\est \emodels \varphi_1 \land \varphi_2$ iff $\est \emodels \varphi_1$ and
$\est \emodels \varphi_2$ iff (by induction hypothesis) $\sest \semodels
\varphi_1$ and $\sest \semodels \varphi_2$ iff (since $\varphi_1 \land \varphi_2
\notin \Fcs$) $\sest \semodels \varphi_1 \land \varphi_2$.

\medskip\noindent%
``$\Leftarrow$'': Since $\Fcs \subseteq \bcl(\Fcs)$, the assumption implies
$\est \emodels \varphi \iff \sest \semodels \varphi \iff \varphi \in \sest$ for
all $\varphi \in \Fcs$ and all $\est \in \ests$; hence, $\ests \seequiv \sest$.
\end{proof}

\CorSEEquivBCL*
\begin{proof}
Let $\befrm \in \bcl(\Fcs)$.  Since $\ests \neq \emptyset$, there is a $\est \in
\ests$ and it holds that $\est \emodels \befrm$ if, and only if, $\sest
\semodels \befrm$ if, and only if, $\est' \emodels \befrm$ for all $\est' \in
\ests$ by \cref{lem:seequiv-bcl}.
\end{proof}

\LemWLPESts*
\begin{proof}
Let first $\ests \emodels \wlp{\eact}{\efrm}$ hold, \ie, $\est \emodels
\wlp{\eact}{\efrm}$ for all $\est \in \ests$, or, equivalently, $\est
\not\emodels \eapre(\eact)$ or $\est \eupd \eact \emodels \efrm$.  Let $\est'
\in \ests \eupd \eact$, \ie, $\est' = \est \eupd \eact$ with $\est \in \ests$
and $\est \emodels \eapre(\eact)$.  Then $\est \eupd \eact \emodels \efrm$.~---
Let now conversely $\ests \eupd \eact \emodels \efrm$ hold, \ie, $\est \eupd
\eact \emodels \efrm$ for all $\est \in \ests$ with $\est \emodels
\eapre(\eact)$.  Let $\est \in \ests$.  If $\est \not\emodels \eapre(\eact)$,
then $\est \emodels \wlp{\eact}{\efrm}$; if $\est \emodels \eapre(\eact)$, then
$\est \eupd \eact \emodels \efrm$ and thus again $\est \emodels
\wlp{\eact}{\efrm}$.
\end{proof}

\LemEUpdSEUpd*
\begin{proof}
Let $\efrm \in \EFrm$.  Then 
\begin{gather*}
\ests \eupd \eact \emodels \efrm
\stackrel{\text{\cref{lem:wlp-ests}}}{\iff}
\ests \emodels \wlp{\eact}{\efrm}
\iff{}\\\qquad
\text{f.\,a.\ }\erepr \in \Wlp{\eact}{\efrm}\text{: }\ests \emodels \erepr
\stackrel{\text{$\eact$ repr., \cref{cor:seequiv-bcl}}}{\iff}{}\\\qquad
\text{ex.\ }\erepr \in \Wlp{\eact}{\efrm} \cap \bcl(\Fcs)\text{ s.\,t.\ }\sest \semodels \erepr
\iff
\efrm \in \sest \eupd \eact
\ \text{.}
\end{gather*}
\end{proof}

\LemECActSem*
\begin{proof}
\cref{it:lem:ecact-sem:zig}~Let $(\ests, \ests') \in \esem{\ecact}$.  Then there
is a $\eact \in \ecact$ such that $\ests' = \ests \eupd \eact$ and $\ests \emodels \eapre(\eact)$.  By
\cref{lem:seequiv-bcl} and $\eact \in \SEAct$ it holds that there is a $\erepr \in \eaPre(\eact) \cap \bcl(\Fcs)$ such that $\sest
\semodels \erepr$ and hence $(\sest, \sest') \in \sesem{\ecact}$ with
$\sest' = \sest \seupd \eact$. By~\cref{lem:eupd-seupd}, $\ests \eupd \eact
\seequiv \sest \seupd \eact$, \ie, $\ests' \seequiv \sest'$.

\smallskip\noindent%
\cref{it:lem:ecact-sem:zag}~Let $(\sest, \sest') \in \sesem{\ecact}$.  Then
there is a $\eact \in \ecact$ such that $\sest' = \sest \seupd \eact$ and $\sest
\semodels \erepr$ for some $\erepr \in \eaPre(\eact) \cap \bcl(\Fcs)$.  By
\cref{lem:seequiv-bcl}, $\ests \neq \emptyset$, and $\eact \in \SEAct$ it holds
that $\ests \emodels \eapre(\eact)$ and hence $(\ests, \ests') \in
\esem{\ecact}$ with $\ests' = \ests \eupd \eact$.  By~\cref{lem:eupd-seupd},
$\ests \eupd \eact \seequiv \sest \seupd \eact$, \ie, $\ests' \seequiv \sest'$.
\end{proof}

\PropEEActSymBisim*
\begin{proof}
\cref{it:prop:eeactsym-bisim:zig}~Let $(\eens, \ests) \eetrans{\eeactsym}
(\eens', \ests')$ be given.  Then there is a $\eens \eectrans{\eecnds \cln
  \eeactsym} \eens'$ with $\ests \emodels \eecnds$ and $(\ests, \ests') \in
\esem{\eecact(\eeactsym)}$.  Since $\ests \seequiv \sest$, $\ests \neq
\emptyset$, and $\eecnds \in \bcl(\Fcs)$ we have, by~\cref{lem:seequiv-bcl},
$\sest \semodels \eecnds$ and,
by~\cref{lem:ecact-sem}\cref{it:lem:ecact-sem:zig}, there exists a $\sest' \in
\SESt$ such that $(\sest, \sest') \in \sesem{\eecact(\eeactsym)}$ and $\ests'
\seequiv \sest'$.  Therefore $(\eens, \sest) \seetrans{\eeactsym} (\eens',
\sest')$ and $\ests' \seequiv \sest'$.

\smallskip\noindent%
\cref{it:prop:eeactsym-bisim:zag}~Let $(\eens, \sest) \seetrans{\eeactsym}
(\eens',\sest')$ be given.  Then there is a $\eens \eectrans{\beecnds \cln
  \eeactsym} \eens'$ with $\sest \semodels \beecnds$ and $(\sest, \sest') \in
\sesem{\eecact(\eeactsym)}$.  Since $\ests \seequiv \sest$ and $\ests \neq
\emptyset$ we have, by~\cref{lem:seequiv-bcl}, $\ests \emodels \beecnds$ and,
by~\cref{lem:ecact-sem}\cref{it:lem:ecact-sem:zag}, there exists $\ests' \in
\ESts$ such that $(\ests, \ests') \in \esem{\eecact(\eeactsym)}$ and $\ests'
\seequiv \sest'$.  Therefore $(\eens, \ests) \eetrans{\eeactsym} (\eens',
\ests')$ and $\ests' \seequiv \sest'$.
\end{proof}

\PropEEActSem*
\begin{proof}
This follows by structural induction on the form of $\eeact \in \EEAct$, where
$\eeactsym \in \bigcup\EEActSym$ is covered by \cref{prop:eeactsym-bisim},
$\befrm{?}$ with $\befrm \in \bcl(\Fcs)$ by \cref{lem:seequiv-bcl}, and all
other cases follow directly from the induction hypothesis.
\end{proof}

\ThmMain*
\begin{proof}
This follows by structural induction on the form of $\eefrm \in \SEEFrm$, where
$\dlbox{\eeact}{\eefrm}$ is a consequence of \cref{prop:eeact-sem}.
\end{proof}
\end{appendix}

\end{document}